\documentclass{aa}
\usepackage{amsmath}
\usepackage{amssymb}
\usepackage{graphicx}
\usepackage{txfonts}
\usepackage{hyperref}
\usepackage{multirow}
\usepackage{float}
\usepackage{orcidlink}
\usepackage{soul}
\usepackage{xcolor}
\hypersetup{colorlinks = true, allcolors = blue}
\usepackage{CJKutf8}

\setul{0.3ex}{1pt}

\begin{document}

   \title{Search for quasar pairs with \textit{Gaia} astrometric data}
   \subtitle{I. Method and candidates}
   \titlerunning{Search for Quasar Pairs with \textit{Gaia} Astrometric Data. I.}
   \author{
           Qihang Chen (\begin{CJK}{UTF8}{gbsn}{陈启航}\end{CJK}) \inst{\ref{BNU_PA},\ref{BNU_FIAA}} \orcidlink{0009-0006-9345-9639}
           \and
           Liang Jing (\begin{CJK}{UTF8}{gbsn}{荆亮}\end{CJK}) \inst{\ref{BNU_PA},\ref{BNU_FIAA}} \orcidlink{0000-0003-1188-9573}
           \and
           Xingyu Zhu (\begin{CJK}{UTF8}{gbsn}{朱星宇}\end{CJK}) \inst{\ref{BNU_PA},\ref{BNU_FIAA}} \orcidlink{0009-0008-9072-4024}
           \and
           Yue Fang (\begin{CJK}{UTF8}{gbsn}{方越}\end{CJK}) \inst{\ref{HBMZU}} \orcidlink{0000-0003-1878-9428}
           \and \\
           Zizhao He (\begin{CJK}{UTF8}{gbsn}{何紫朝}\end{CJK}) \inst{\ref{NCU_DOP},\ref{NCU_CRAHEP},\ref{PMO}} \orcidlink{0000-0001-8554-9163}
           \and
           Zhuojun Deng (\begin{CJK}{UTF8}{gbsn}{邓卓君}\end{CJK}) \inst{\ref{BNU_PA},\ref{BNU_FIAA}} \orcidlink{0009-0008-8080-3124}
           \and
           Cheng Xiang (\begin{CJK}{UTF8}{gbsn}{向成}\end{CJK}) \inst{\ref{BNU_PA},\ref{BNU_FIAA}}
           \orcidlink{0009-0000-6610-8979}
           \and
           Jianghua Wu (\begin{CJK}{UTF8}{gbsn}{吴江华}\end{CJK}) \inst{\ref{BNU_PA},\ref{BNU_FIAA}} \thanks{Corresponding author: \href{mailto:jhwu@bnu.edu.cn}{jhwu@bnu.edu.cn}} \orcidlink{0000-0002-8709-6759}
           }
   \institute{
              School of Physics and Astronomy, Beijing Normal University, Beijing, 100875, China \\ \email{jhwu@bnu.edu.cn} \label{BNU_PA}
              \and
              Institute for Frontier in Astronomy and Astrophysics, Beijing Normal University, Beijing, 102206, China \label{BNU_FIAA}
              \and
              College of Intelligent Systems Science and Engineering, Hubei Minzu University, Enshi, Hubei, 445000, China \label{HBMZU}
              \and
              Department of Physics, Nanchang University, Nanchang, 330031, China \label{NCU_DOP}
              \and
              Center for Relativistic Astrophysics and High Energy Physics, Nanchang University, Nanchang, 330031, China \label{NCU_CRAHEP}
              \and
              Purple Mountain Observatory, Chinese Academy of Sciences, Nanjing, Jiangsu, 210023, China \label{PMO}
             }
   \date{Received 8 May 2025 / Accepted 8 January 2026}

    \abstract{Quasar pairs, a special subclass of galaxy pairs, are valuable in the investigation of quasar interaction and clustering, the coevolution between the two quasars' host galaxies, and the growth of supermassive black holes, as well as the formation and evolution of galaxies overall. However, quasar pairs on kiloparsec scales are observationally rare. The scarcity of available samples hindered the deeper exploration and statistics of these objects. In this work, we apply an astrometric method to systematically search for quasar candidates within a transverse distance of 100 kpc to known quasars in the Million Quasar Catalog. These candidates are \textit{Gaia} sources with proper motions and parallaxes that are consistent with zero. A visual inspection of the sample was performed to remove the contamination of crowded stellar fields and nearby galaxies. A total of 4\,112 quasar pair candidates were isolated, with a median member separation of 8.81{\arcsec}, a median \textit{Gaia} $G$-band magnitude of 20.52, and a median redshift of 1.61. Following a comparison of our catalog with three major candidate quasar pair catalogs, we identified 3\,984 new quasar pair candidates that had previously been overlooked in the three catalogs. Several interesting quasar pair candidates are highlighted and discussed in this work. We also offer a brief discussion of our quasar selection and several techniques for improving the success rate of quasar pair selection. Extensive spectroscopic follow-up campaigns are currently underway to validate their astrophysical nature.}
    \maketitle

\section{Introduction} \label{sec1}
Quasars are a class of extremely luminous active galactic nuclei (AGNs), each hosting a supermassive black hole (SMBH) surrounded by an accretion disk from which powerful radiation originates. In the unified model of AGN, an extensive dusty torus surrounds the central engine and the broad-line regions (BLRs), with the BLRs situated on both sides of the accretion disk. The narrow-line regions (NLRs) extend beyond the torus, enveloping it at larger radial distances \citep[e.g.,][]{Antonucci1993AGNunification, Urry1995RadioAGNunification, Netzer2015AGNunification}. Since the discovery of quasars in the early 1960s \citep{Schmidt1963qso1st3C273, Sandage1965qso1st}, numerous surveys have discovered close to a million objects of this kind \citep[the Million Quasar Catalog,][]{Flesch2023MQCv8}. The most recently released DESI DR1 even reported 1.65 million quasars, substantially extending the known quasar population \citep{DESICollaboration2025DR1}.

If two quasars or galaxies encounter and interact, they may form a type of system known as ``quasar pair,'' which makes up only a minute fraction of the general quasar population. Observationally, quasar pairs exhibit typical separations of less than several arcseconds and nearly identical redshifts, corresponding to sub-milliparsec (sub-Mpc) transverse physical distances. According to \citet{DeRosa2019QPreview}, quasar pairs can be classified into two subclasses, dual quasar and binary quasar, based on the transverse distances between pair members. The dual quasar is defined as an interacting galaxy system containing two active nuclei that are not mutually gravitationally bound. The transverse distance between the two quasars ranges from $\sim$ 1 pc to $\sim$ 100 kpc. The binary quasar refers to two quasars that are gravitationally bound and forming a Keplerian binary. The two quasars are separated by a parsec (pc) to sub-pc transverse distance. Quasar pairs are natural results of the hierarchical evolution of galaxies. They are important laboratories to study the quasar interaction, coevolution between the two hosts, environment overdensity, clustering, large-scale structure, the growth of SMBHs, and the formation and evolution of galaxies in extreme environments as well as the early Universe \citep[e.g.,][]{Begelman1980BAGN, Impey1998QPLSS, YuQJ2002MNRASBSMBHevolution, Hennawi2006BQinSDSSclustering, Myers2008BQclustering, LiuX2011AGNpairfraction, ShenY2010BQclustering, ShenY2023QPfraction, Merritt2013AGNevolution, Sandrinelli2014QPenvironment, Sandrinelli2018QPenvironment, Richarte2025QPLSS, HouMC2020LiuXAGNpairXstats, AnT2022BSMBHVLBI, XuWC2024DAGNVLBA, WangJM2023BSMBHfinalparsec}. They are also potential origins of nanohertz gravitational waves (nHz-GWs) in the early Universe \citep{Kelley2019aQPnHzGW, ChenYF2020BSMBHevolutionGW, ShenY2023DSMBHnHzGW}.

However, the rarity of confirmed quasar pairs inhibited deep exploration and statistical investigations. According to recent statistical work, only $\sim$ 160 quasar pairs with a transverse distance of less than 100 kpc are publicly confirmed to date \citep[Big MAC DR1,][]{Pfeifle2025TheBigMACDR1}. It is therefore important to search for more of them. The search for quasar pairs can be dated back to \citet{Stockton1972firstQP} and \citet{Wampler1973secondQP}. However, their discoveries are not true quasar pairs but projected cases. The first populations of quasar pairs were discovered by some optical and radio surveys \citep[e.g.,][]{Bolton1976QP, Meylan1987QP, Djorgovski1987BQ, Munoz1998BQMGC2214+3550}. With the successive launches of large-scale imaging sky surveys such as SDSS \citep{York2000SDSS}, WISE \citep{Wright2010WISE}, Pan-STARRS \citep{Chambers2016PanSTARRS}, and recent DESI Legacy Imaging Surveys \citep[DESI-LS,][]{Dey2019DESILS}, a significant number of quasar pairs and their candidates have been discovered by utilizing a color-based method and spectroscopic follow-up \citep[e.g.,][]{Hennawi2010HighZ24BQ, Green2010BQJ1254+0846, ChenYC2021aDoubleQinDECamLS}. Several tens of quasar pairs have also been discovered as by-products of the search for strongly lensed quasars \citep[e.g.,][]{YueMH2023HighZLeQQP, Dawes2023MultiLeQBQinDESILS, HeZZ2025LeQDQPQ}, for example, the Binary Quasar sub-catalog compiled from the Gravitationally lensed quasars in \textit{Gaia} project\footnote{\url{https://research.ast.cam.ac.uk/lensedquasars/binaries.html}} \citep[GLQG,][]{Lemon2017GLQG-I-method}. Furthermore, a recently proposed image decomposition method has also discovered a few quasar pairs \citep[e.g.,][]{Silverman2020DSMBHinHSCSSP, TangSL2021DQinHSCSSP}. Combining optical imaging and spectroscopic surveys seems to be a promising and efficient approach to searching for quasar pairs.

In addition, several indirect methods have been used to search for quasar pairs with various characteristics, such as periodic variability \citep{Sillanpaa1988OJ287BSMBH, Graham2015aBSMBHperiodicity, Graham2015bBSMBHvariability, LiaoWT2021BSMBHperiodicity, Davis2024BSMBHvariability}, radial velocity drift \citep{GuoHX2019BSMBHrvd, ZhengQ2023BQrvd}, double-peaked emission lines \citep{ZhouHY2004BQdpbel, Boroson2009BSMBHdpbel, TangSM2009BSMBHdpbel, ShenY2010BSMBHdpbel, Kim2020RedAGNdpbel, Kim2020DAGNdpbel, ZhengQ2024DAGNdpbel}, asymmetric emission line profiles \citep{Eracleous2012aBSMBHasyELP}, and continuum anomalies \citep{YanCS2015BSMBHcontinuum}, as well as the observations of binary active cores in radio or X-ray images \citep{LiuFK2003BSMBHinRaido, Komossa2003BAGNinChandraX}. However, these indirect methods are not entirely reliable. The periodic variability induced by orbital motion can be easily imitated by the intrinsic irregular variability of the quasars \citep{Charisi2016BSMBHperiodicity}. Accretion disks can also produce double-peaked emission line profiles \citep{Eracleous1997ADdpbel, Eracleous2012bADdpbel}, and some double-peaked narrow-line objects have been confirmed to be single sources \citep{Foord2020wrongDAGNdpbel}. A single source can also produce velocity drift and a double-peaked profile, while altering the profile of emission lines \citep{Du&Wang2023SpiralArmdpbel}. Methods based on color or the spectral energy distribution (SED) also have their limitations, as the quasar region in the color-color diagram is heavily contaminated by some specific spectral types of stars \citep{Ross2012SDSS-IIIBOSSDR9Qselection, WuXB2012OptIRQuasarCands, Yeche2020DESIQselection}.

The astrometric search for quasars is one of the most remarkable and promising methods. Quasars are observed as nearly stationary objects on the celestial sphere due to their enormous distance from Earth. Quasar selection with the astrometric method can be traced back to \citet{Koo1986firstAstrometry}. They separated quasars from white dwarfs with the assistance of a proper motion criterion and demonstrated the power of the astrometric method. To make efficient and unbiased searches for quasars, several efforts have drawn on the astrometric method \citep[e.g.,][]{ShuYP2019GaiaunWISEQuasarCat, WuQQ2023AstrometryMIRQuasarCands}. For instance, \citet{Heintz2015pmplx, Heintz2018SNRpm, Heintz2020SNRpmplx} applied a pure astrometric method to select quasars and derived a conservative selection efficiency of $\sim 75\,\%$, indicating great success in quasar selection with astrometry. \citet{Lemon2019GLQG-IIIPMSIG} presented a stricter criterion of proper motion significance (PMSIG) to assist the search for lensed quasars. Similarly, a probability-based zero-proper motion method ($f_{\textsl{PM0}}$), which is more complete in a computational sense than PMSIG, was proposed to help select quasar candidates behind the Galactic plane \citep[GPQ,][]{FuYM2021GPQ-I}. The probability-based zero-proper motion method was also used in \citet{FuYM2024CatNorth} and \citet{FuYM2025CatSouth} to construct a whole sky candidate quasar catalog. These efforts validate the capability of astrometry in quasar selection.

For the quasar pair search, the joint significance of the total proper motion $\chi^2$ successfully selected some quasar pairs with quite small member separations around 1{\arcsec} \citep{Souchay2022DQlpm, Makarov2022DQlpm}. On the other hand, some novel methods have also discovered several tens of spatially unresolved or marginally resolved dual AGNs, quasars, and SMBHs that are separated by less than 1{\arcsec} (from sub-kpc to a few kpc). These discoveries open up new avenues to in-depth exploration of the "final parsec" problem, which refers to the theoretical difficulty of bringing two SMBHs from a parsec-scale separation to a distance where gravitational wave emission can efficiently drive them to final coalescence \citep[e.g.,][]{Milosavljevic2001BSMBHfinalparsec, Merritt2005BSMBHfinalparsec, ZhangF2023BIMBHfinalparsec, Bromley2024BSMBHfinalparsec, Koo2024BSMBHfinalparsec}. These novel methods include anomalous astrometry \citep{Makarov2012Quasometry, Makarov2023GaiaWISEDQLeQ, WuQQ2022SQUAB-I, JiX2023SQUAB-II}, optical flickering or photometric centroid shift \citep[e.g.,][]{LiuY2015BAGNCentroidShift-I, LiuY2016BAGNCentroidShift-II}, \textit{Gaia} multipeak \citep[GMP,][]{Mannucci2022GMP, Mannucci2023GMP, Ciurlo2023GMP, Scialpi2024GMP, WuQQ2024GMP-DULAG}, and astrometric jitter and disturbance techniques \citep[Vastrometry,][]{ShenY2019VODKA, ShenY2021AstrometryHighZDQ, Uppal2024AstrometricJitter, HwangHC2020VODKAmethodology}. Vastrometry was initially used in an optical-band project titled Vastrometry for Off-nucleus and Dual sub-Kpc AGN \citep[VODKA, e.g.,][]{ChenYC2022VODKAHSTDQ, ChenYC2023VODKAVLBAsearchDQ, ChenYC2025VODKAGeminiHST} and has recently been applied to radio bands within the framework of Varstrometry for Dual AGN using Radio interferometry \citep[VaDAR, e.g.,][]{WangHC2023VODKAradio, Schwartzman2024VaDAR}. In contrast to the aforementioned small-separation pairs, a serendipitously discovered quasar pair at \textit{z} = 1.76 and with a member separation of 8.76{\arcsec} indicates that the selection of point sources without significant proper motion next to known quasars could be an effective way to search for quasar pairs across a wider range of member separations \citep{Altamura2020SerendipitousBQpm}.

Referring to the probability-based astrometric criteria from \citet{FuYM2021GPQ-I} and inspired by \citet{Altamura2020SerendipitousBQpm}, we initiated a systematic search for new quasars located in the vicinity of known quasars to find new quasar pairs with a combination of \textit{Gaia} astrometric data and the Million Quasar Catalog (MQC). This paper details that search and is organized as follows. The data exploited in this work, method details, and the derived candidates are introduced in Section \ref{sec2}. The basic statistical and color properties of the quasar pair candidates, the comparative advantages of our sample over three major candidate quasar pair catalogs, and several interesting candidates are described in Section \ref{sec3}. The discussion on our selection and future improvement is given in Section \ref{sec4}. We summarise our conclusions in Section \ref{sec5}. Throughout this paper, we adopt a flat lambda cold dark matter ($\Lambda$CDM) cosmology with $\Omega_{\Lambda}$ = 0.7, $\Omega_M$ = 0.3, and $H_0$ = 70 km $\cdot$ s$^{-1}$ $\cdot$ Mpc$^{-1}$.

\section{Data and method} \label{sec2}
The catalogs used in this work include version 8.0 of MQC and \textit{Gaia} Data Release 3 (\textit{Gaia} DR3). The imaging data of DESI-LS and Pan-STARRS were used to visually inspect and refine the astrometrically selected quasar pair candidates.

\subsection{Data} \label{sec2.1}
The MQC builds on the Half Million Quasar Catalog \citep{Flesch2015HMQC}. The sources included in this catalog can be traced back to the first release (version 0.1) on April 22, 2009. After 14 years of continuous expansion and the addition of quasar samples confirmed from large numbers of multiband photometric and spectroscopic surveys, the final version 8.0 of the MQC (hereafter, MQCv8)\footnote{MILLIQUAS v8.0 (2023) update, \url{https://heasarc.gsfc.nasa.gov/W3Browse/all/milliquas.html}} was published on August 2, 2023 \citep{Flesch2023MQCv8}. The MQCv8 contains 1\,021\,800 quasars and high-probability candidates, 60.7\,\% of which possess \textit{Gaia} DR3 astrometry. As the SDSS DR16Q \citep{Lyke2020SDSSDR16Q} constitutes the primary source of this catalog, the resulting large and robust quasar sample provides a sufficient basis for a systematic search for quasar pairs.

\textit{Gaia} is a powerful space astrometric satellite launched by the European Space Agency (ESA) in December 2013. The mission is for a space-based, astrometric, and photometric all-sky survey at optical wavelengths, providing more than one billion accurate measurements of astrometric parameters for a variety of objects \citep{deBruijne2012GaiaMission, GaiaCollaboration2016GaiaMission}. Under the \textit{Gaia}'s unprecedented measurement accuracy of a few microarcseconds, extragalactic sources and objects in the Milky Way can be distinguished effectively by astrometric parameters such as proper motion and parallax. \textit{Gaia} published its DR3 on June 13, 2022  \citep{GaiaCollaboration2023DR3},  containing 1.8 billion astrometric objects and 1.5 billion sources with measured proper motions and parallaxes, with an accuracy as high as 0.001 mas/yr.

\subsection{Method} \label{sec2.2}
Based on the combination of the latest MQC and the most enormous astrometric database, searching for extragalactic sources with proper motion and parallax results consistent with zero near the spectroscopically confirmed quasars has immense advantages in the catalog volume, astrometric accuracy, and the quasar selection bias. The following three major steps are executed to obtain the quasar pair candidates:

1) cross-match between MQCv8 and \textit{Gaia} DR3;

2) astrometric filtering on the cross-matched catalog;

3)  visual inspection of the astrometrically selected pair candidates.

\subsubsection{Cross-match} \label{sec2.2.1}

The catalog cross-matching tool TOPCAT \citep{Taylor2005TOPCAT} was employed to carry out the first step of our method. To start with, quasars in MQCv8 were searched for in the SIMBAD database\footnote{\url{http://simbad.cds.unistra.fr/simbad/}} \citep{Wenger2000SIMBAD} using a 1{\arcsec} radius to update their redshift. To search for higher-redshift quasar pair candidates and extend the existing quasar pair catalog, quasars with z > 0.5 were selected as the base sample, which includes 908\,990 quasars.  We then cross-matched this quasar base sample with \textit{Gaia} DR3 using a search radius of 1{\arcsec} to exclude quasars with imprecise radio or X-band positions. The \textit{Gaia} match closest to the position of the quasar was considered as its \textit{Gaia} counterpart. This yields a \textit{Gaia} optical quasar sample, which consists of 582\,725 quasars with \textit{Gaia} detections.

Based on the updated redshift, the angular separation corresponding to a transverse distance of 100 kpc (angular diameter distance) to each optical quasar was calculated under the flat $\Lambda$CDM cosmology. These separations were used to perform the cross-match of \textit{Gaia} DR3 sources to the optical quasars with \textit{Gaia} positions. A total of 105\,971 \textit{Gaia} sources within 100 kpc to the MQCv8 quasars were selected to constitute a preliminary catalog, marked as \texttt{100kpcCAT}. Then, the \texttt{100kpcCAT} was searched again in SIMBAD within a radius of 1{\arcsec} to exclude those spectroscopically confirmed sources. In addition, a magnitude cut of $G \leqslant 21$ mag was performed to ensure the reliability of the astrometric measurements \citep{GaiaCollaboration2023DR3}. Finally, a sample of 85\,839 sources with proper motion and parallax measurements but without redshift was isolated as the input sample of the next step of the astrometric filtering. For convenience, we named this sample \texttt{astrometricCAT}.

\subsubsection{Astrometric filtering} \label{sec2.2.2}
To support the astrometric filtering, reference samples of spectroscopically confirmed quasars and stars are required. SDSS, LAMOST, and DESI provide extensive spectral libraries of quasars and stars, making them ideal for reference sample construction. We retrieved all quasars and stars from SDSS DR18 \citep{Almeida2023SDSSDR18} via its Science Archive Server\footnote{\url{https://dr18.sdss.org/optical/spectrum/search}} (SAS). For LAMOST, quasars and stars were extracted from its DR12 v1.0 general catalog\footnote{\url{http://www.lamost.org/dr12/v1.0/}} \citep{CuiXQ2012LAMOST}. For DESI, quasars and stars were obtained from its recent DR1\footnote{\url{https://data.desi.lbl.gov/doc/access/}}. All the sources were cross-matched with \textit{Gaia} DR3 to extract sources that have both proper motion and parallax measurements.

Given the distinct Galactic coordinate distributions of quasars and stars, we randomly selected \textit{Gaia} cross-matched quasars and stars within each of six Galactic latitude bins: 0{\degr} $\leqslant \lvert b \rvert <$ 20{\degr}, 20{\degr} $\leqslant \lvert b \rvert <$ 30{\degr}, 30{\degr} $\leqslant \lvert b \rvert <$ 45{\degr}, 45{\degr} $\leqslant \lvert b \rvert <$ 60{\degr}, 60{\degr} $\leqslant \lvert b \rvert <$ 75{\degr}, and 75{\degr} $\leqslant \lvert b \rvert \leqslant$ 90{\degr}. Each sky region bin contains equal numbers of quasars and stars. The selection generated a reference sample based on the sky region, which contains 704\,942 quasars and the same number of stars. It is denoted as \texttt{Good\_Sky} for convenience and the quasar and star subsamples are denoted as \texttt{GoodQSO\_Sky} and \texttt{GoodSTAR\_Sky}, respectively.

Considering that the astrometric accuracy in \textit{Gaia} DR3 is strongly dependent on source brightness \citep{GaiaCollaboration2023DR3}, another reference sample that is related to the \textit{Gaia} $G$-band magnitude is required. To exclude extremely faint sources that may have poor astrometric accuracy and to ensure consistency with the range of the \texttt{astrometricCAT}'s \textit{Gaia} $G$-band magnitude, we randomly sampled \textit{Gaia} cross-matched quasars and stars according to three \textit{Gaia} $G$-band magnitude bins: 12 $\leqslant G <$ 19, 19 $\leqslant G <$ 20, and 20 $\leqslant G \leqslant$ 21. Each magnitude bin contains equal numbers of quasars and stars. Moreover, the random sampling was designed to ensure identical $G$-band magnitude distributions between quasars and stars. This design can eliminate the impact of magnitude discrepancy on astrometric filtering. The sampling generated a reference sample based on the magnitude, which contains 697\,779 quasars and the same number of stars. Similarly, it is denoted as \texttt{Good\_Mag} and the quasar and star subsamples are denoted as \texttt{GoodQSO\_Mag} and \texttt{GoodSTAR\_Mag}, respectively.

\begin{figure}
    \centering
    \includegraphics[width=0.48\textwidth]{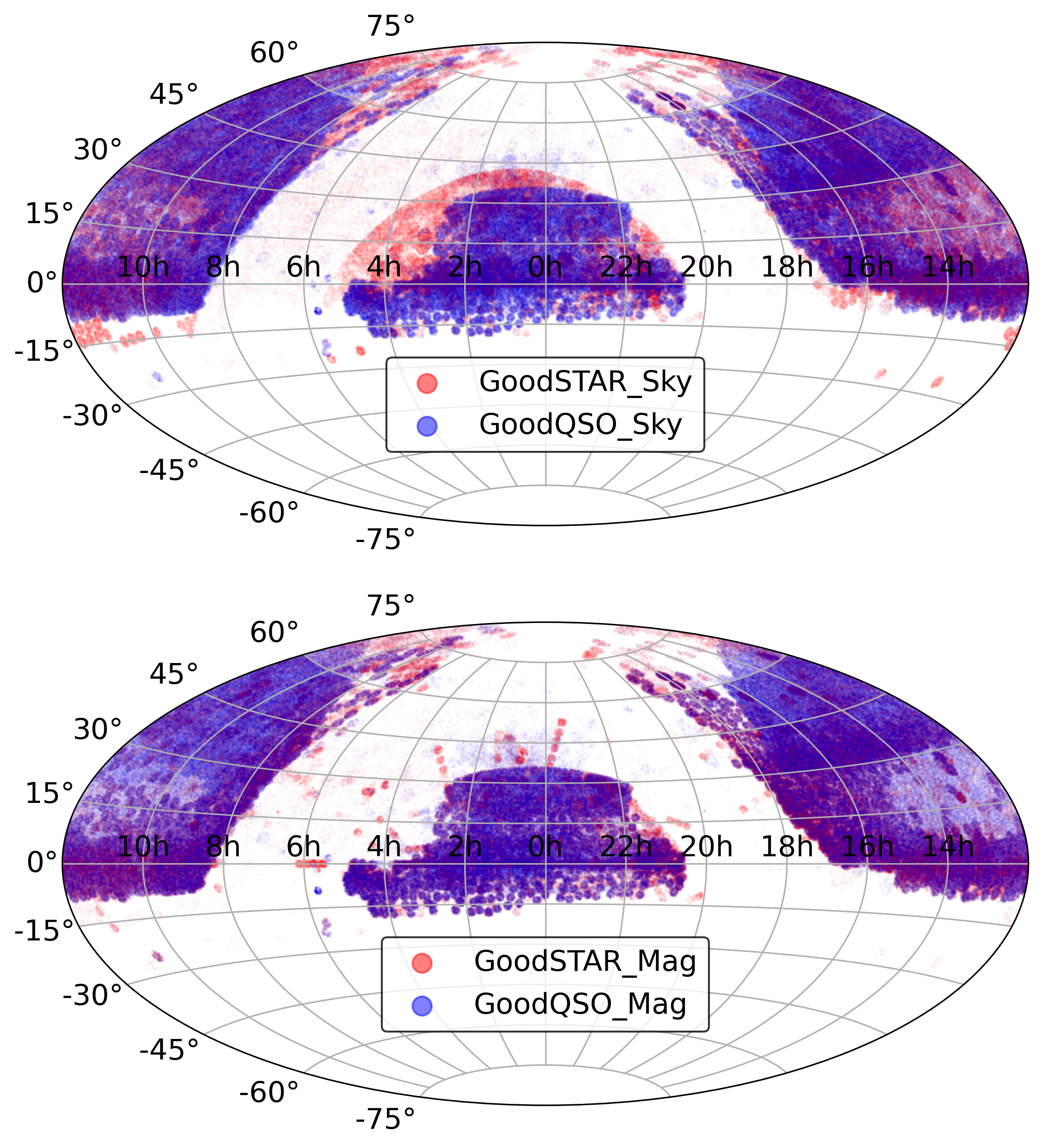}
    \caption{Hammer-Aitoff equatorial projection of sky region binned (top panel) and \textit{Gaia} $G$-band magnitude binned (bottom panel) reference samples. The blue scatters in the top and bottom panels represent the \texttt{GoodQSO\_Sky} and \texttt{GoodQSO\_Mag}, respectively. The red scatters in the top and bottom panels represent the \texttt{GoodSTAR\_Sky} and \texttt{GoodSTAR\_Mag}, respectively.}
    \label{refCAT_Aitoff}
\end{figure}

The distributions of the \texttt{Good\_Sky} and \texttt{Good\_Mag} in equatorial coordinates are displayed in Figure \ref{refCAT_Aitoff}. The equal volume of the sky region binned or magnitude binned quasar and star subsamples ensures an unbiased astrometric filtering. The two reference samples are the keys to the astrometric filtering.

Utilizing several astrometric measurements provided by \textit{Gaia} DR3, two types of astrometric criteria are proposed to select extragalactic candidates efficiently. The astrometric measurements include the proper motion in right ascension and declination (\texttt{pmRA} and \texttt{pmDE}), proper-motion uncertainties in the two directions (\texttt{e\_pmRA} and \texttt{e\_pmDE}), the correlation coefficient between \texttt{pmRA} and \texttt{pmDE} (\texttt{pmRApmDEcor}), parallax (\texttt{Plx}), and parallax uncertainties (\texttt{e\_Plx}). Following \citet{FuYM2021GPQ-I}, the probability density of zero proper motion is employed to set the astrometric criteria. Equation 10 in \citet{FuYM2021GPQ-I} defines the probability density of zero proper motion of a \textit{Gaia} DR2 source, but several of its parameters are no longer applicable to \textit{Gaia} DR3. Therefore, we adopted Equation 7 in \citet{FuYM2024CatNorth} to calculate the probability density of zero proper motion of a \textit{Gaia} DR3 source,
\begin{align}
    \nonumber
    f_{\textsl{0pm}} =& \frac{1}{2\pi\sigma_{\mu_{\alpha^*}}\sigma_{\mu_{\delta}}\sqrt{1-\rho}} \\ &\times \textsl{exp}\left\{-\frac{1}{2(1-\rho^2)} \left[\left(\frac{\mu_{\alpha^*}}{\sigma_{\mu_{\alpha^*}}}\right)^2-\frac{2\rho \mu_{\alpha^*} \mu_{\delta}}{\sigma_{\mu_{\alpha^*}}\sigma_{\mu_{\delta}}}+\left(\frac{\mu_{\delta}}{\sigma_{\mu_{\delta}}}\right)^2\right]\right\},
\end{align}
\label{Eq1}where $\mu_{\alpha^*}$ is proper motion in right ascension, $\mu_{\delta}$ is proper motion in declination, $\rho$ is the correlation coefficient between $\mu_{\alpha^*}$ and $\mu_{\delta}$, and $\sigma_{\mu_{\alpha^*}}$, and $\sigma_{\mu_{\delta}}$ are the proper-motion uncertainties. According to the equation, sources with smaller proper motions will have larger $f_{\textsl{0pm}}$ under the same uncertainty level.

Considering that the parallax of the nearby Galactic stars is generally significant, whereas extragalactic sources exhibit almost zero parallaxes, an additional criterion is introduced, namely, the probability density of zero parallaxes, to eliminate the Galactic stars with coincidental near-zero proper motion. Similarly, the probability density of zero parallax of a \textit{Gaia} DR3 source is defined as
\begin{equation}
    f_{\textsl{0plx}} = \frac{1}{\sqrt{2\pi}\sigma_\varpi}\cdot\textsl{exp}\left[-\frac{1}{\ 2\ } \cdot \left(\frac{\varpi}{\sigma_\varpi}\right)^2\right],
\end{equation}
\label{Eq2}where $\varpi$ and $\sigma_\varpi$ are the parallax and its uncertainty, respectively. According to the definition, sources with smaller parallaxes will have larger $f_{\textsl{0plx}}$ under the same uncertainty level.

\begin{figure*}
    \centering
    \includegraphics[width=0.9\textwidth]{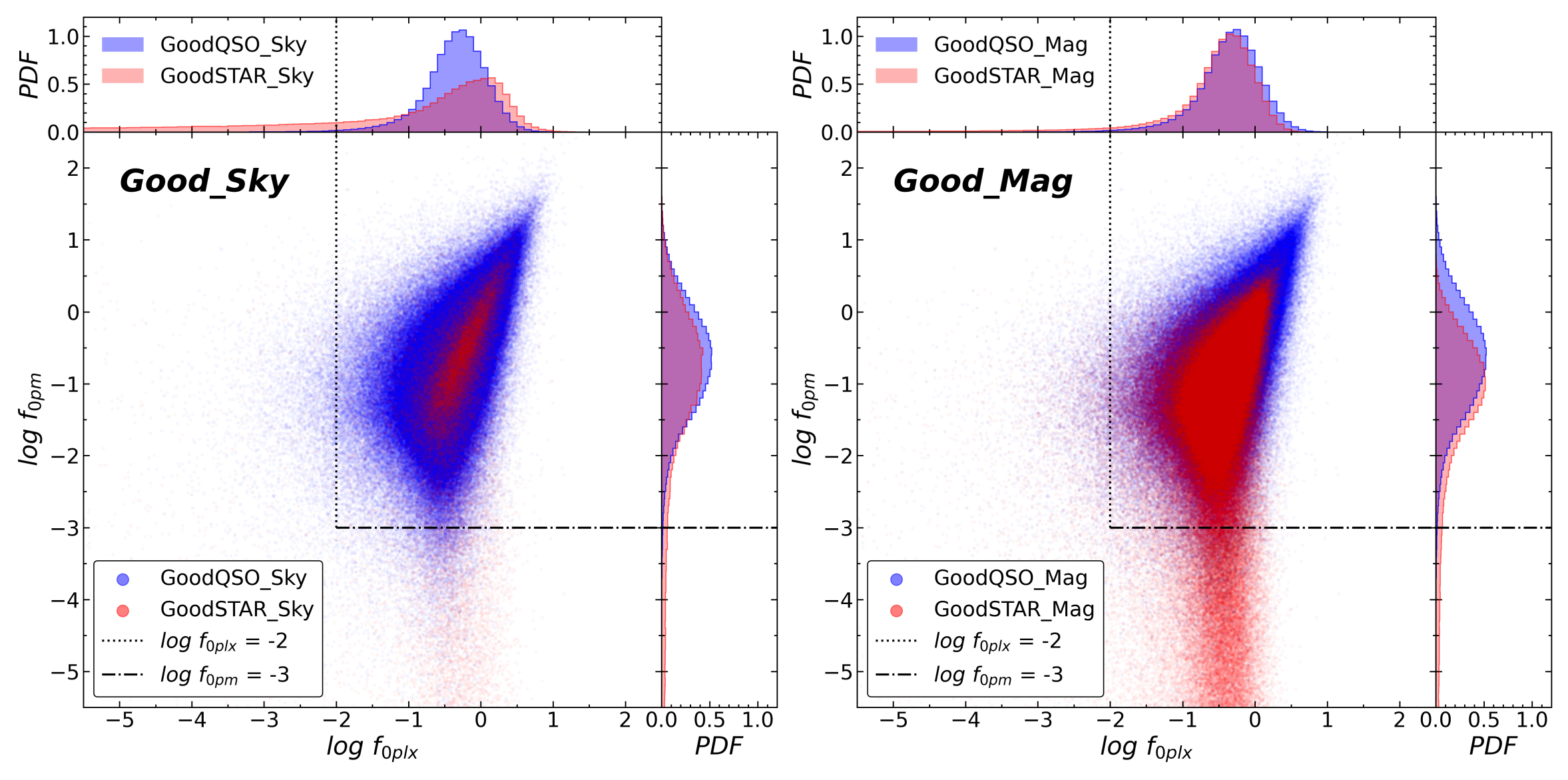}
    \caption{log $f_{\textsl{0pm}}$ vs. log $f_{\textsl{0plx}}$ distributions of the \texttt{Good\_Sky} (left panel) and \texttt{Good\_Mag} (right panel), along with the quasar selection regions of the adopted astrometric criteria. The left panel shows the distributions of the sky-unbinned GoodQSO\_Sky (blue points) and GoodSTAR\_Sky (red points), while the right panel displays the corresponding distributions for the magnitude-unbinned GoodQSO\_Mag (blue) and GoodSTAR\_Mag (red). The black dotted dashed lines represent the $f_{\textsl{0pm}}$ criteria, while the dotted ones represent the $f_{\textsl{0plx}}$ criteria. All the histograms are normalized as probability density distributions (PDFs). Due to the long-tailed distributions of stars in log $f_{\textsl{0pm}}$ and log $f_{\textsl{0plx}}$, i.e., many stars fall outside the plot range, both subplots display only small portions of the entire distributions.}
    \label{am_cria_skymag_unbinned}
\end{figure*}

The $f_{\textsl{0pm}}$ and $f_{\textsl{0plx}}$ values were calculated for all sources in the \texttt{Good\_Sky}, \texttt{Good\_Mag}, and \texttt{astrometricCAT}. The logarithm of them (log $f_{\textsl{0pm}}$ and log $f_{\textsl{0plx}}$) were adopted for better visualization. Figure \ref{am_cria_skymag_unbinned} displays the log $f_{\textsl{0pm}}$ versus log $f_{\textsl{0plx}}$ distributions of the unbinned \texttt{Good\_Sky} (left panel) and \texttt{Good\_Mag} (right panel). Both distributions employ optimally determined thresholds of criteria through iterative attempts; specifically, log $f_{\textsl{0pm}}\geqslant-3$ and log $f_{\textsl{0plx}}\geqslant-2$. From the distribution of the sky region unbinned \texttt{Good\_Sky}, it appears that this set of criteria effectively filters out the majority of stars. In contrast, the distribution of the magnitude unbinned \texttt{Good\_Mag} indicates that a significant number of stars still cannot be efficiently removed.

\begin{figure*}
    \centering
    \includegraphics[width=\textwidth]{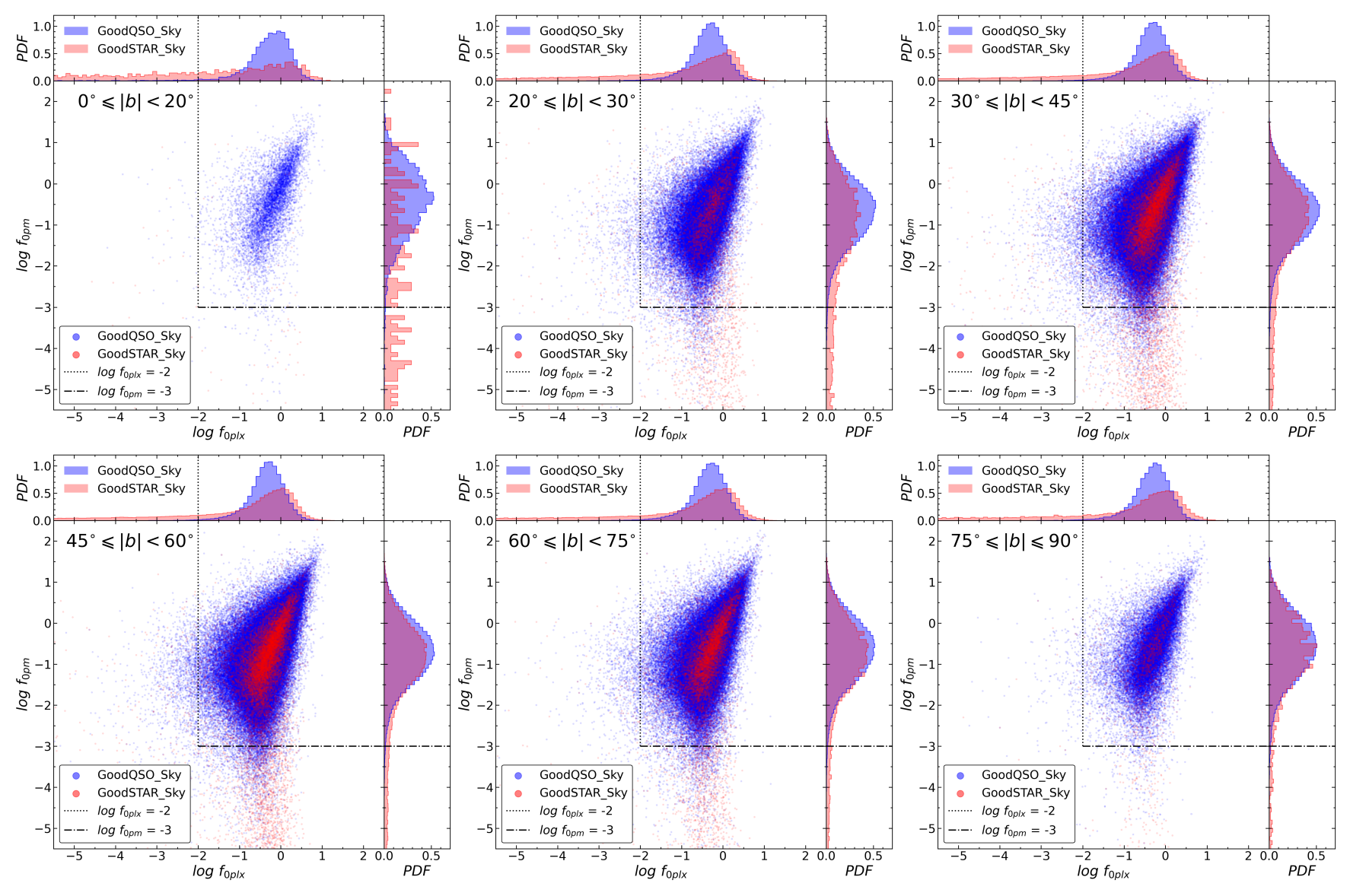}
    \caption{Same as the left panel of Figure \ref{am_cria_skymag_unbinned}, but for the \texttt{Good\_Sky} at different Galactic latitude bins.}
    \label{am_cria_skybinned}
\end{figure*}

To further examine whether the quasar selection efficiency is affected by sky region, the distributions of aforementioned six sky binned \texttt{Good\_Sky} are plotted in Figure \ref{am_cria_skybinned}, along with the optimal thresholds of criteria, which were iteratively determined. Table \ref{Tab_amcira_Sky} lists the quasar selection completeness, purity, and stellar contamination for each sky bin under the corresponding astrometric criteria. It is evident that when $\lvert b \rvert \geqslant$ 20{\degr}, the completeness, purity, and stellar contamination of quasar selection remain stable at approximately 98\,\%, 95\,\%, and 5\,\%, respectively, with increasing Galactic latitude. The determined thresholds of the criteria are identical in each sky bin, indicating that the quasar selection efficiency is almost unaffected by sky region, at least within the context of this work. This result markedly differs from the conclusions of \citet[][hereafter, Heintz20]{Heintz2020SNRpmplx}, where quasar selection purity increases from $\sim$ 20\,\% to $\sim$ 80\,\% with increasing Galactic latitude. The underlying reasons are explained and discussed in Section \ref{sec4.1}.

\begin{table*}
    \caption{Astrometric criteria based on the Galactic latitude ($b$) and their corresponding completeness, purity, and stellar contamination.}
    \centering
    \begin{tabular}{c|c|c|c|c|l}
    \hline \hline
    Galactic latitude & Astrometric criteria & Completeness & Purity & Stellar contamination & Source number in bin\\
    \hline
    \ \ 0{\degr} $\leqslant \lvert b \rvert <$ 20{\degr} &  & 84.35\,\% & 99.22\,\% & 0.78\,\% & $N_Q$ = $N_S$ = 8\,901\\
    20{\degr} $\leqslant \lvert b \rvert <$ 30{\degr} & \multirow{4}{*}{log $f_{\textsl{0pm}}\ \geqslant\ -3$} & 97.83\,\% & 96.07\,\% & 3.93\,\% & $N_Q$ = $N_S$ = 93\,119\\
    30{\degr} $\leqslant \lvert b \rvert <$ 45{\degr} & \multirow{4}{*}{log $f_{\textsl{0plx}}\ \geqslant\ -2$} & 98.07\,\% & 94.88\,\% & 5.12\,\% & $N_Q$ = $N_S$ = 198\,068\\
    45{\degr} $\leqslant \lvert b \rvert <$ 60{\degr} &  & 98.12\,\% & 94.51\,\% & 5.49\,\% & $N_Q$ = $N_S$ = 226\,976\\
    60{\degr} $\leqslant \lvert b \rvert <$ 75{\degr} &  & 98.16\,\% & 94.21\,\% & 5.79\,\% & $N_Q$ = $N_S$ = 143\,273\\
    75{\degr} $\leqslant \lvert b \rvert \leqslant$ 90{\degr} &  & 98.05\,\% & 94.58\,\% & 5.42\,\% & $N_Q$ = $N_S$ = 34\,605\\
    Total &  & 97.90\,\% & 94.81\,\% & 5.19\,\% & $N_Q$ = $N_S$ = 704\,942\\
    \hline
    \end{tabular}
    \tablefoot{The Galactic latitude range of 0{\degr} $\leqslant \lvert b \rvert <$ 20{\degr} is the Galactic plane. Due to the lack of confirmed quasars within it and the abnormal purity and stellar contamination, the final astrometric filtering on the \texttt{astrometricCAT} sources within this region was not performed. The last column lists the numbers of quasars ($N_Q$) and stars ($N_S$) in each sky region bin prior to astrometric filtering.}
    \label{Tab_amcira_Sky}
\end{table*}

In the 0{\degr} $\leqslant \lvert b \rvert <$ 20{\degr} sky bin, the purity and stellar contamination exhibit anomalies. This region corresponds to the Galactic plane, where the available quasar reference sample is quite small, containing only 8\,901 quasars. To ensure a balanced comparison, the star sample was limited to the same size, resulting in a total sample that was too small for robust statistics and a reliable determination of the quasar selection efficiency. This issue similarly affects the 20{\degr} $\leqslant \lvert b \rvert <$ 30{\degr} and 75{\degr} $\leqslant \lvert b \rvert \leqslant$ 90{\degr} bins (see the last column in Table \ref{Tab_amcira_Sky}). Furthermore, this area suffers from significant stellar contamination, along with obscuration, extinction, and reddening caused by Galactic dust and gas, which substantially degrade the astrometric accuracy of sources. Quasar selection based purely on astrometric filtering is unsuitable for such crowded stellar fields \citep[CSFs, e.g.,][]{FuYM2021GPQ-I}. Therefore, before applying astrometric filtering to the \texttt{astrometricCAT}, we removed candidates located in the Galactic plane. Similarly, pure astrometric filtering is also inapplicable in the Large and Small Magellanic Clouds \citep[LMC and SMC, e.g.,][]{FuYM2025CatSouth} and candidates within these CSFs were also excluded prior to the astrometric filtering.

\begin{figure}
    \centering
    \includegraphics[width=0.48\textwidth]{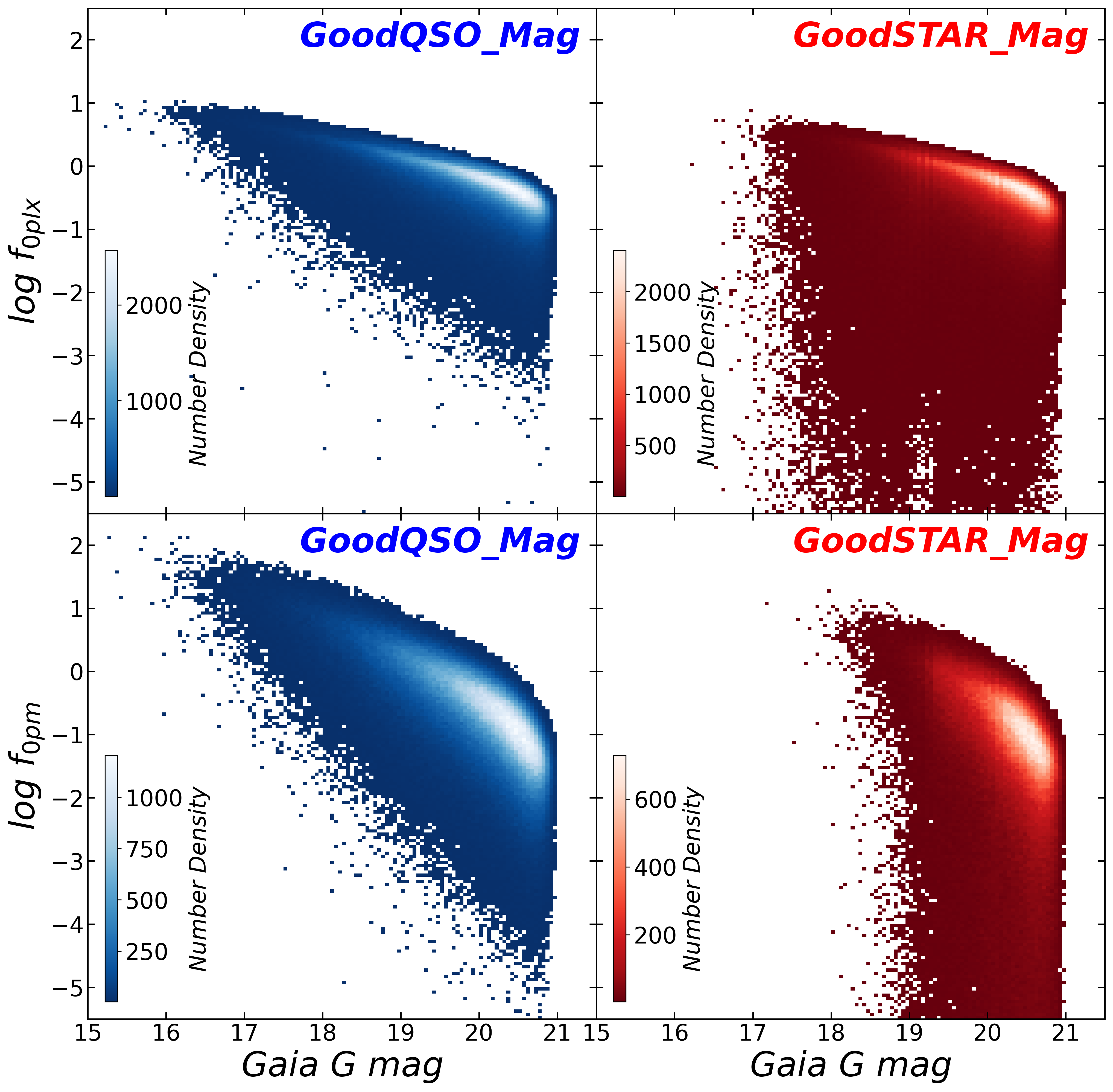}
    \caption{ log $f_{\textsl{0plx}}$ vs. \textit{Gaia} $G$-band magnitude (top row) and log $f_{\textsl{0pm}}$ vs. \textit{Gaia} $G$-band magnitude (bottom row) distributions of the \texttt{GoodQSO\_Mag} (blue) and \texttt{GoodSTAR\_Mag} (red). The color bar in each subplot represents the number density of the distribution. Note: the color map in each subplot follows its own color bar, and thus, identical colors across different subplots do not represent equivalent number densities. All the subplots display only small parts of the whole distributions to exhibit the critical data region.}
    \label{mag_am_relation}
\end{figure}

The overall quasar selection completeness, purity, and stellar contamination based on Galactic latitude-dependent astrometric criteria are 97.90\,\%, 94.81\,\%, and 5.19\,\%, respectively, showing minimal variation across different sky regions. However, these results remain overly optimistic and unreliable because source brightness also significantly impacts astrometric quasar selection. Specifically, source brightness significantly affects the astrometric accuracy in \textit{Gaia}, particularly the uncertainties of proper motion and parallax. Therefore,  log $f_{\textsl{0plx}}$ and log $f_{\textsl{0pm}}$ versus \textit{Gaia} $G$-band magnitude are plotted in Figure \ref{mag_am_relation} to demonstrate how the magnitude influences the two probability densities. The two subplots in the left column display the two probability densities versus $G$-band magnitude distributions of \texttt{GoodQSO\_Mag}, while the two subplots in the right column show the corresponding distributions of \texttt{GoodSTAR\_Mag}. Evidently, both probability densities of \texttt{GoodQSO\_Mag} and \texttt{GoodSTAR\_Mag} decrease with increasing $G$-band magnitude, which is equivalent to the behavior shown in Figure 2 of \citet{GaiaCollaboration2023DR3}, where the uncertainties of proper motion and parallax increase with increasing $G$-band magnitude.

\begin{figure*}
    \centering
    \includegraphics[width=\textwidth]{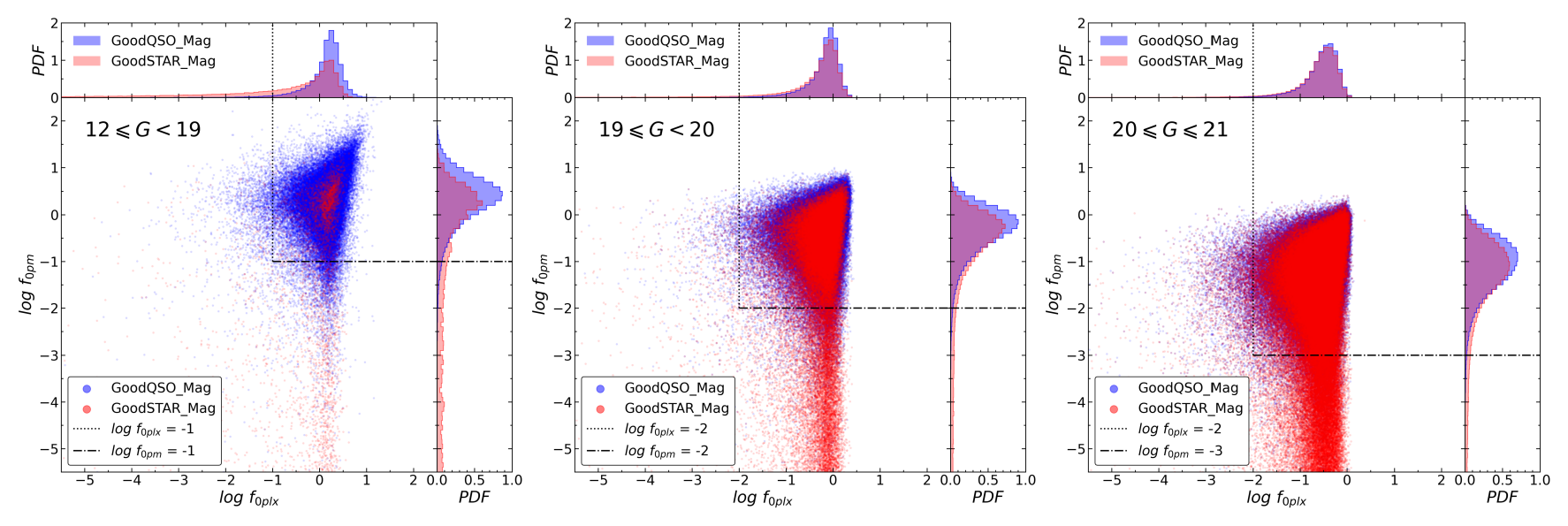}
    \caption{Similar to Figure \ref{am_cria_skybinned}, but for the \texttt{Good\_Mag} at different \textit{Gaia} $G$-band magnitude bins.}
    \label{am_cria_magbinned}
\end{figure*}

Based on \textit{Gaia} $G$-band magnitude bins, the new astrometric criteria employ a multithreshold strategy to minimize stellar contamination. Figure \ref{am_cria_magbinned} displays the log $f_{\textsl{0pm}}$ versus log $f_{\textsl{0plx}}$ distributions of the \texttt{Good\_Mag} at different \textit{Gaia} $G$-band magnitude bins, along with the multithreshold quasar selection regions. All these distributions employ optimally determined thresholds of criteria through iterative attempts. It is evident that the \textit{Gaia} $G$-band magnitude significantly affects the quasar selection efficiency. Therefore, instead of applying uniform astrometric criteria across the entire magnitude range, we determined optimal astrometric criteria for sources in different magnitude bins. Table \ref{Tab_amcira_Gmag} lists the completeness, purity, and stellar contamination for each magnitude bin under the corresponding astrometric criteria. While maintaining relatively high completeness ($\textgreater 90\,\%$), the purity decreases with increasing $G$-band magnitude, whereas the stellar contamination shows an opposite trend. This demonstrates that faint sources significantly degrade the purity of quasar selection when using pure astrometric filtering, particularly for $G \geqslant 19$ sources (the last two subplots in Figure \ref{am_cria_magbinned}).

\begin{table*}
    \caption{Astrometric criteria based on the \textit{Gaia} $G$-band magnitude ($G$) and their corresponding completeness, purity, and stellar contamination.}
    \centering
    \begin{tabular}{c|c|c|c|c|l}
    \hline \hline
    Magnitude range & Astrometric criteria & Completeness & Purity & Stellar contamination & Source number in bin\\
    \hline
    \multirow{2}{*}{$12 \leqslant G < 19$} & log $f_{\textsl{0pm}}\ \geqslant\ -1$ & \multirow{2}{*}{91.33\,\%} & \multirow{2}{*}{96.71\,\%} & \multirow{2}{*}{3.29\,\%} & \multirow{2}{*}{$N_Q$ = $N_S$ = 86\,553}\\
    & log $f_{\textsl{0plx}}\ \geqslant\ -1$ &  &  & \\
    \hline
    \multirow{2}{*}{$19 \leqslant G < 20$} & log $f_{\textsl{0pm}}\ \geqslant\ -2$ & \multirow{2}{*}{97.50\,\%} & \multirow{2}{*}{77.10\,\%} & \multirow{2}{*}{22.90\,\%} & \multirow{2}{*}{$N_Q$ = $N_S$ = 233\,821}\\
    & log $f_{\textsl{0plx}}\ \geqslant\ -2$ &  &  & \\
    \hline
    \multirow{2}{*}{$20 \leqslant G \leqslant 21$} & log $f_{\textsl{0pm}}\ \geqslant\ -3$ & \multirow{2}{*}{97.65\,\%} & \multirow{2}{*}{61.80\,\%} & \multirow{2}{*}{38.20\,\%} & \multirow{2}{*}{$N_Q$ = $N_S$ = 377\,405}\\
    & log $f_{\textsl{0plx}}\ \geqslant\ -2$ &  &  & \\
    \hline
    $12 \leqslant G \leqslant 21$ & log $f_{\textsl{0pm}}\ \geqslant\ -3$ & \multirow{2}{*}{97.91\,\%} & \multirow{2}{*}{69.24\,\%} & \multirow{2}{*}{30.76\,\%} & \multirow{2}{*}{$N_Q$ = $N_S$ = 697\,779}\\
    (Unbinned) & log $f_{\textsl{0plx}}\ \geqslant\ -2$ &  &  & \\
    \hline
    Total & Criteria of the first three bins & 96.82\,\% & 69.37\,\% & 30.63\,\% & $N_Q$ = $N_S$ = 697\,779\\
    \hline
    \end{tabular}
    \tablefoot{The fourth row displays selection results without binning the $G$ magnitude. The last row is the comprehensive selection performance of the final multi-threshold astrometric criteria. The numbers of quasars ($N_Q$) and stars ($N_S$) before astrometric filtering in each magnitude bin are given in the last column.}
    \label{Tab_amcira_Gmag}
\end{table*}

The adopted multithreshold astrometric criteria were defined as follows:
\begin{enumerate}[1)]
    \item for $12 \leqslant G < 19$: $\log f_{\textsl{0pm}} \geqslant -1$ and $\log f_{\textsl{0plx}} \geqslant -1$;
    \item for $19 \leqslant G < 20$: $\log f_{\textsl{0pm}} \geqslant -2$ and $\log f_{\textsl{0plx}} \geqslant -2$;
    \item for $20 \leqslant G \leqslant 21$: $\log f_{\textsl{0pm}} \geqslant -3$ and $\log f_{\textsl{0plx}} \geqslant -2$.
\end{enumerate}
Applying these criteria yielded comprehensive quasar selection completeness, purity, and stellar contamination of 96.82\,\%, 69.37\,\%, and 30.63\,\%, respectively. It should be noted that the quasar selection efficiency remains slightly optimistic, as several factors may further degrade the performance. Related discussions are presented in Section \ref{sec4.1}. The final multithreshold filtering was performed on 74\,357 \texttt{astrometricCAT} sources outside three CSFs (the Galactic plane, LMC, and SMC), and isolated a sample of 4\,286 quasar candidates, denoted as \texttt{MGQPCpreCAT}.

\subsubsection{Visual inspection} \label{sec2.2.3}
The astrometric filtering isolates extragalactic sources along with a subset of stars exhibiting near-zero proper motions and parallaxes, yet it fails to distinguish galaxies from quasars. Therefore,  \texttt{MGQPCpreCAT} still requires morphological inspection to exclude extended sources.

DESI-LS DR10 provides morphological classifications for all observed sources, including parameters denoted as PSF (point sources) and SER (Sersic surface brightness profiles). The \texttt{MGQPCpreCAT} was cross-matched with DR10 (\texttt{ls\_dr10.tractor} catalog\footnote{\url{https://datalab.noirlab.edu/data-explorer?showTable=ls_dr10.tractor}}) using a 1{\arcsec} search radius to obtain morphological parameters of the candidates. The cross-matching yielded 2\,821 candidates with morphological parameters, the vast majority of which are classified as PSF. Although three major CSFs had been removed prior to the astrometric filtering in Section \ref{sec2.2.2}, a small number of candidates located in Galactic stellar clusters and spiral arms of nearby galaxies remained. Fortunately, these candidates are distinctly different from others and can be easily removed through visual inspection.

The $g$-, $r$-, and $z$-band imaging data of the \texttt{MGQPCpreCAT} sources were obtained from DESI-LS DR10\footnote{\url{https://www.legacysurvey.org/viewer}} to support the visual inspection. Although DR10 covers nearly half of the entire sky ($\sim$20\,000 deg$^2$), it lacks imaging data near the Galactic plane and in several specific sky areas. Therefore, the Pan-STARRS $g$-, $r$-, and $z$-band imaging data\footnote{\url{https://ps1images.stsci.edu/cgi-bin/ps1cutouts}} were used to supplement areas where DESI-LS DR10 images are either missing or of poor quality. All pseudo-color images were composited from $g$-, $r$-, and $z$-band images using \texttt{HumVI} \citep{Marshall2015SWHumVI, Marshall2016HumVI} via RGB combination.

\begin{figure*}
    \centering
    \includegraphics[width=\textwidth]{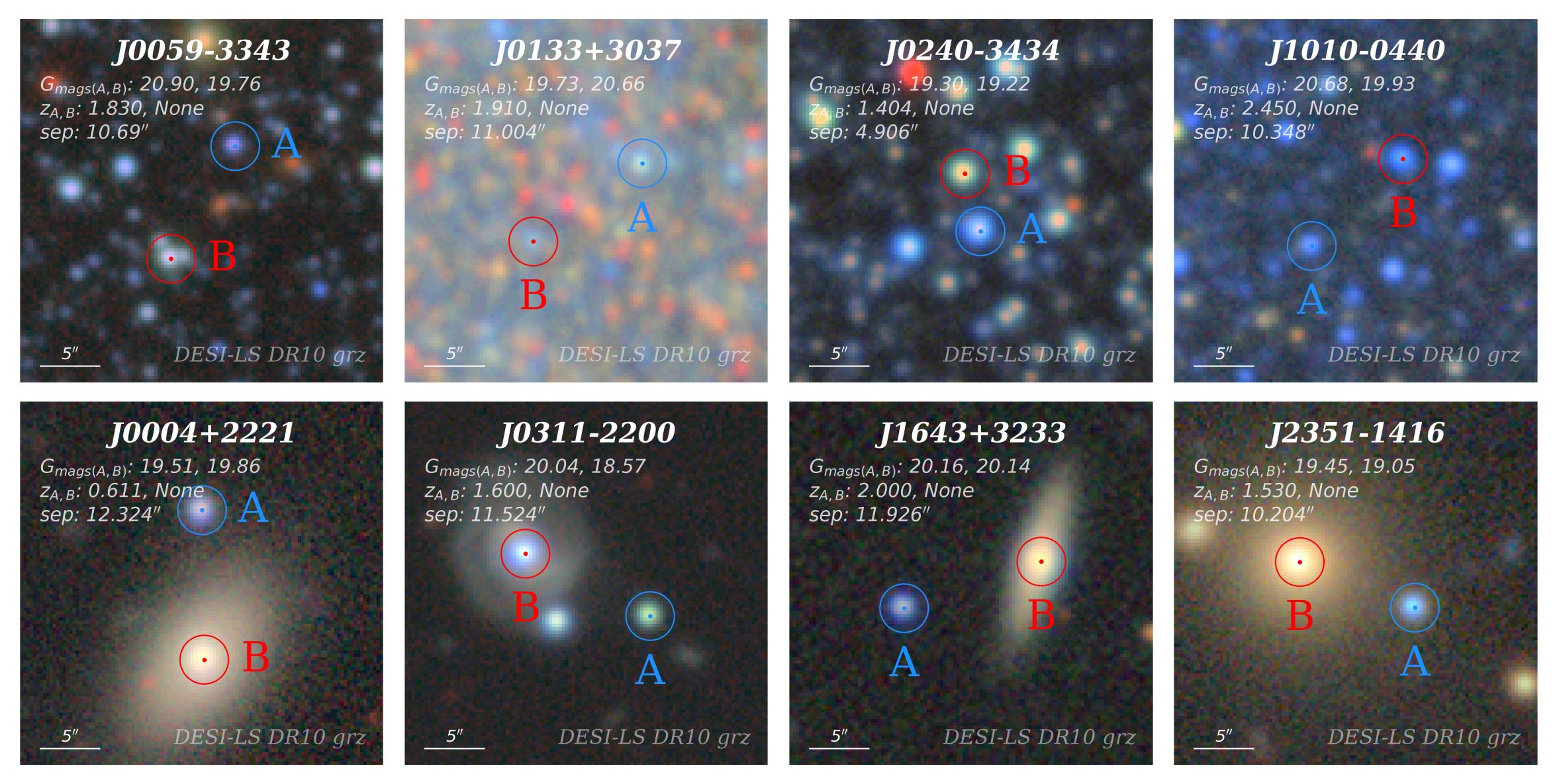}
    \caption{Examples of visually rejected CSFs (top row) and morphologically removed nearby galaxies (bottom row). All these cutout images are obtained from DESI-LS DR10 $grz$ bands and composited using \texttt{HumVI}. The blue and red circles with center points mark the quasars from MQCv8 and the astrometrically selected extragalactic candidates, respectively. Their \textit{Gaia} $G$-band magnitudes, redshifts, and separation between them are denoted by $G_{mags(A,B)}$, ``$z_{A,B}$'', and ``sep'', respectively. All of these sources are unlikely quasar pairs and were removed. The size of each image is 30{\arcsec}\,$\times$\,30{\arcsec.} North is up and east is to the left.}
    \label{DFs_NGs_examples}
\end{figure*}

\begin{figure*}
    \centering
    \includegraphics[width=0.9\textwidth]{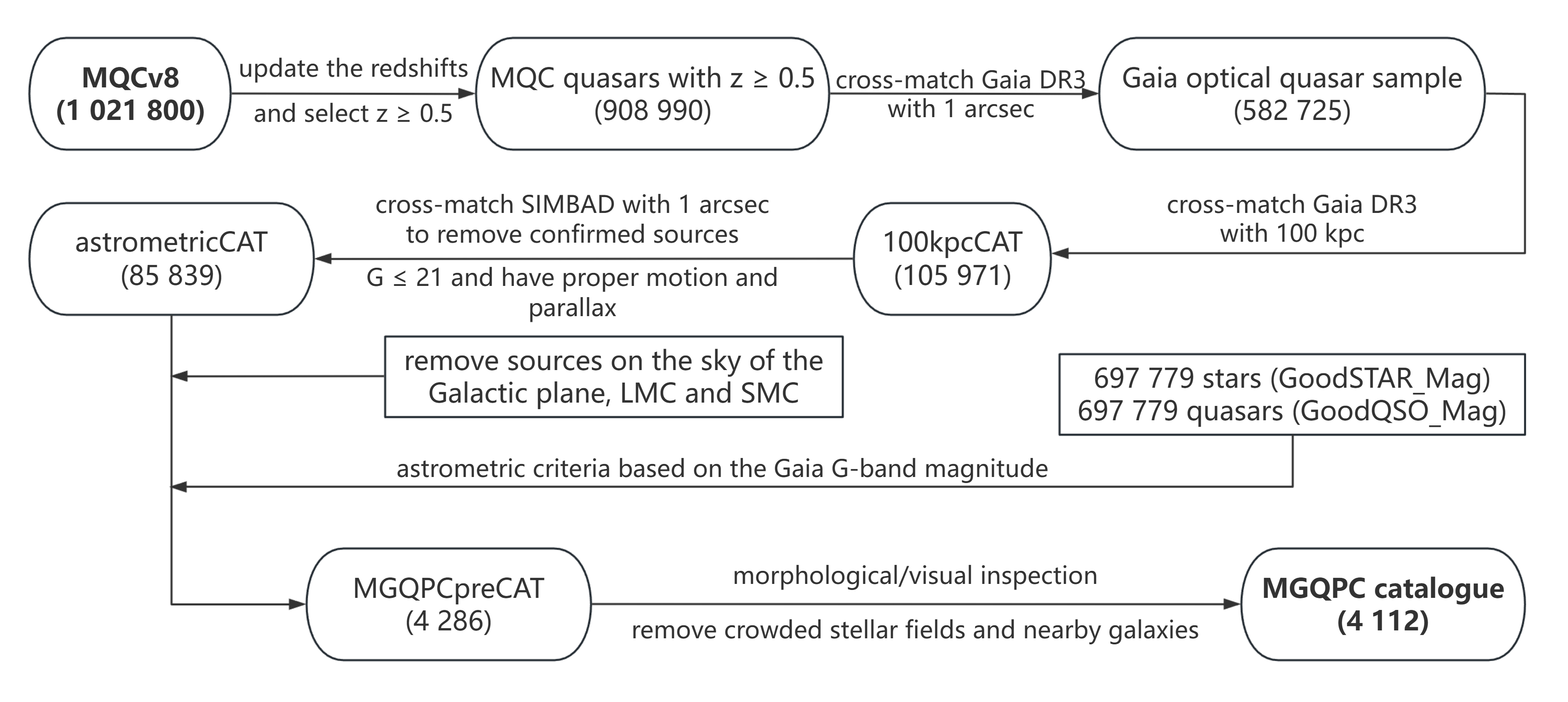}
    \caption{Flowchart of the MGQPC catalog selection process.}
    \label{Flowchart}
\end{figure*}

With the support of morphological parameters and pseudocolor images, the visual inspection was performed on the 2\,821 candidates to exclude nearby galaxies, CSFs, and galaxies misclassified as PSF. The remaining 1\,465 candidates lacking morphological parameters were also visually inspected. Altogether, 174 CSFs and nearby galaxies were removed from the \texttt{MGQPCpreCAT}, with examples shown in Figure \ref{DFs_NGs_examples}. A total of 4\,112 plausible quasar pair candidates were selected. We designate them as MGQPCs, which stands for ``quasar pair candidates derived from the cross-match of MQC and Gaia''. The flowchart of the MGQPCs selection is shown in Figure \ref{Flowchart}.

\section{Overview of the MGQPC catalog} \label{sec3}
The basic statistical features of the final sample of quasar pair candidates (MGQPCs) and the advantages of the MGQPC catalog are described and validated in this section. Several fascinating pair systems are also described.

\subsection{Basic features of MGQPCs} \label{sec3.1}
\begin{figure*}
    \centering
    \includegraphics[width=\textwidth]{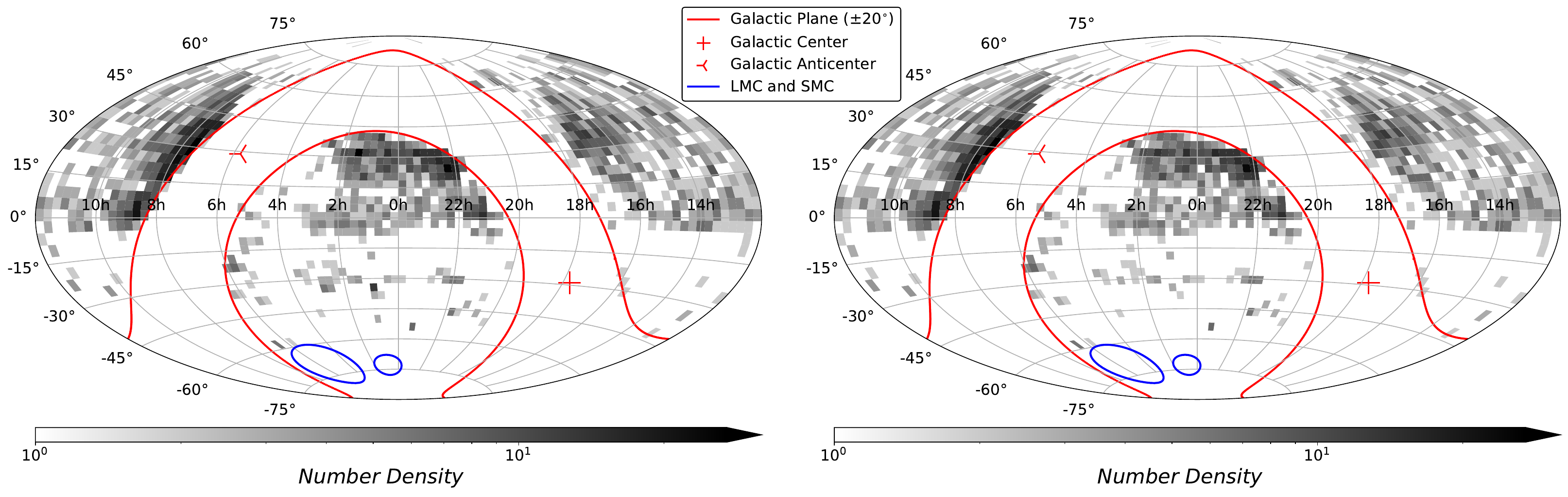}
    \caption{Hammer-Aitoff projection of the \texttt{MGQPCpreCAT} (left panel) and the visually selected MGQPCs (right panel). On both panels, the red lines roughly demarcate the Galactic plane (0{\degr} $\leqslant \lvert b \rvert <$ 20{\degr}), while the red cross and Y-shaped point mark the Galactic center and anti-center, respectively. The coverages of the LMC and SMC are bounded by the blue ellipses on both panels.}
    \label{aitoff_map}
\end{figure*}

\begin{figure}
    \centering
    \includegraphics[width=0.48\textwidth]{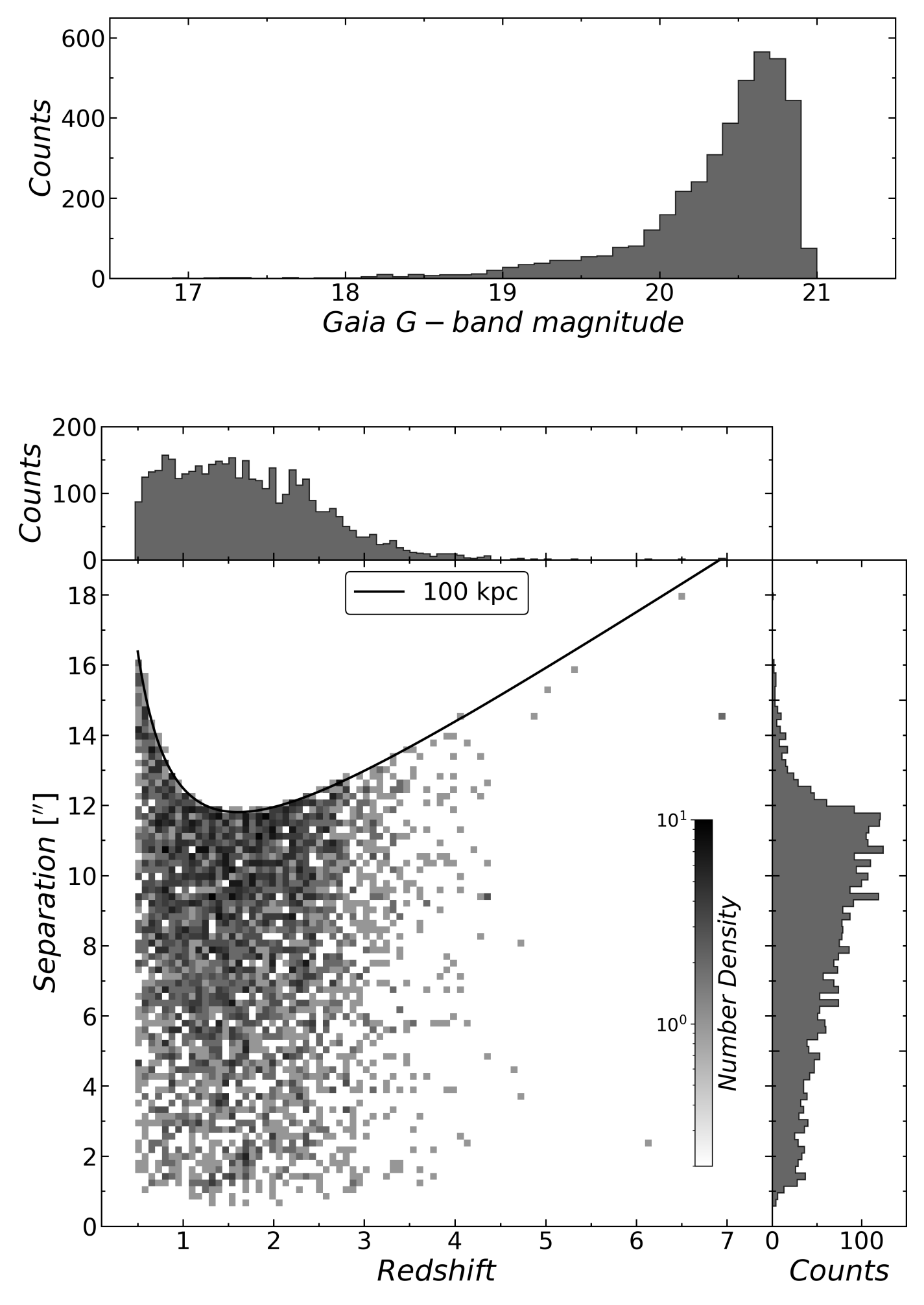}
    \caption{Top panel: \textit{Gaia} $G$-band magnitude histogram of the MGQPCs. Bottom panel: Separation v.s. redshift (the known quasar) of the MGQPCs. The solid line represents the 100 kpc criterion mentioned in Section \ref{sec2.2.1}.}
    \label{MGQPC_statistics}
\end{figure}

\begin{figure*}
    \centering
    \includegraphics[width=\textwidth]{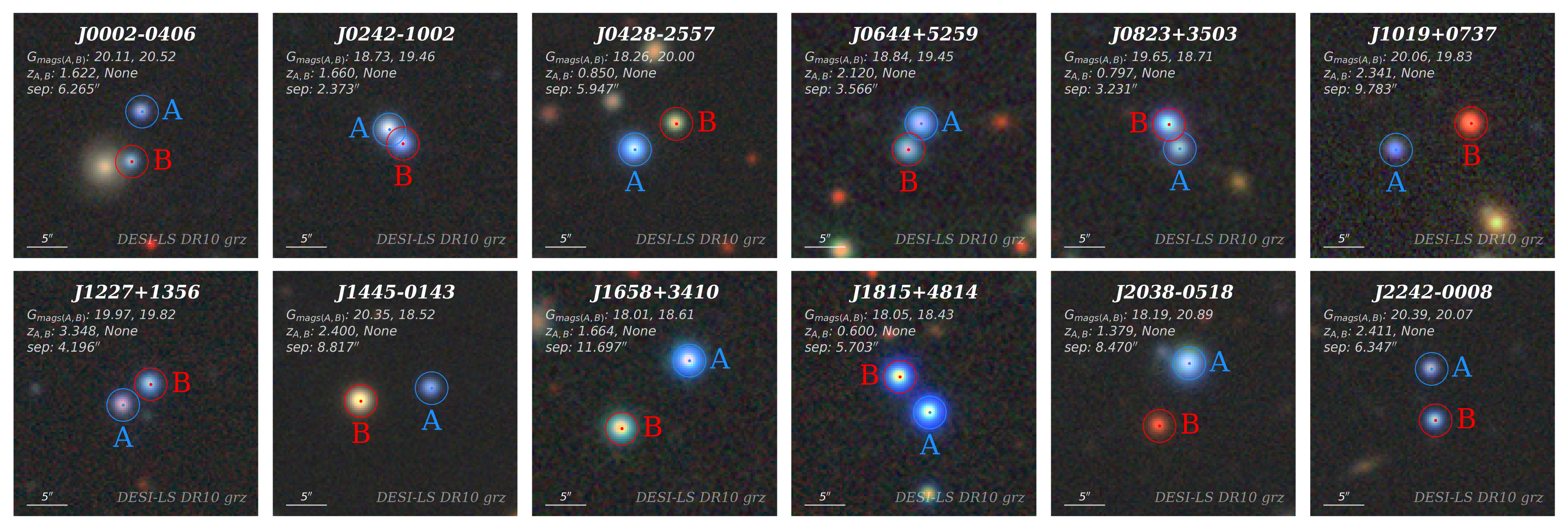}
    \caption{Examples of MGQPCs. All these cutout images are obtained from DESI-LS DR10 $grz$ bands and composited using \texttt{HumVI}. The marks of blue and red circles, text annotations, image size, and direction are the same as in Figure \ref{DFs_NGs_examples}.}
    \label{MGQPC_examples}
\end{figure*}

The comparison of the equatorial projection of the \texttt{MGQPCpreCAT} (left panel) and the visually selected MGQPCs (right panel) is displayed in Figure \ref{aitoff_map}. Most MGQPCs are distributed in the northern hemisphere and visual inspection only induces minor variations in the number densities of several small regions. Figure \ref{MGQPC_statistics} plots the MGQPCs' \textit{Gaia} $G$-band magnitude histogram (top panel) and distribution of the separation versus redshift (bottom panel). The redshifts are invoked from the known member quasars. Faint sources and moderately large separations dominate the MGQPC catalog, with a median \textit{Gaia} $G$-band magnitude of 20.52 and a median member separation of 8.81{\arcsec}. The rapid decrease in the $G$-band magnitude histogram at $G \gtrsim 20.8$ mag is primarily due to missing or less precise astrometry at faint magnitudes \citep[e.g., Section 3.1 and Figure 2 in][]{GaiaCollaboration2021EDR3}. The number of MGQPCs decreases with increasing redshift and decreasing separation. Examples of MGQPCs are displayed in Figure \ref{MGQPC_examples}. A description of the primary columns in the MGQPC catalog is given in Table \ref{Tab_MGQPC_columns}.

\begin{table*}
    \caption{Description of the primary columns in the MGQPC catalog.}
    \centering
    \begin{tabular}{l|c|c}
    \hline \hline
    Column(s) & Description & Example value (unit) \\
    \hline
    SystemName & The pair name designated by the center coordinates & J000004.061-041259.407 \\
    DR3Name\_A, DR3Name\_B & Unique \textit{Gaia} DR3 identifier & Gaia DR3 2447741272610542720 \\
    RA\_A, RA\_B & Right ascension (J2000.0) & 0.018109143 (degree) \\
    DEC\_A, DEC\_B & Declination (J2000.0) & -4.215430503 (degree) \\
    Z\_A & Redshift of the known quasar & 1.851 \\
    f\_0plx\_A, f\_0plx\_B & Probability density of zero parallax & 0.303 \\
    f\_0pm\_A, f\_0pm\_B & Probability density of zero proper motion & 0.282 \\
    Gmag\_A, Gmag\_B & Gaia $G$-band magnitude & 20.252 (mag) \\
    BPmag\_A, BPmag\_B & Gaia $BP$-band magnitude & 20.337 (mag) \\
    RPmag\_A, RPmag\_B & Gaia $RP$-band magnitude & 19.967 (mag) \\
    sep\_AB & Angular separation between the two members in the pair & 11.449 ({\arcsec}) \\
    seplim & Cross-matching radius & 11.863 ({\arcsec}) \\
    \hline
    \end{tabular}
    \tablefoot{The columns with suffix ``\_A'' describe the known quasar, while those with suffix ``\_B'' describe the nearby quasar candidate. The last entry ``seplim'' represents the separation corresponding to a transverse distance of 100 kpc at the known quasar redshift (see Section \ref{sec2.2.1} for details).}
    \label{Tab_MGQPC_columns}
\end{table*}

\begin{figure*}
    \centering
    \includegraphics[width=0.9\textwidth]{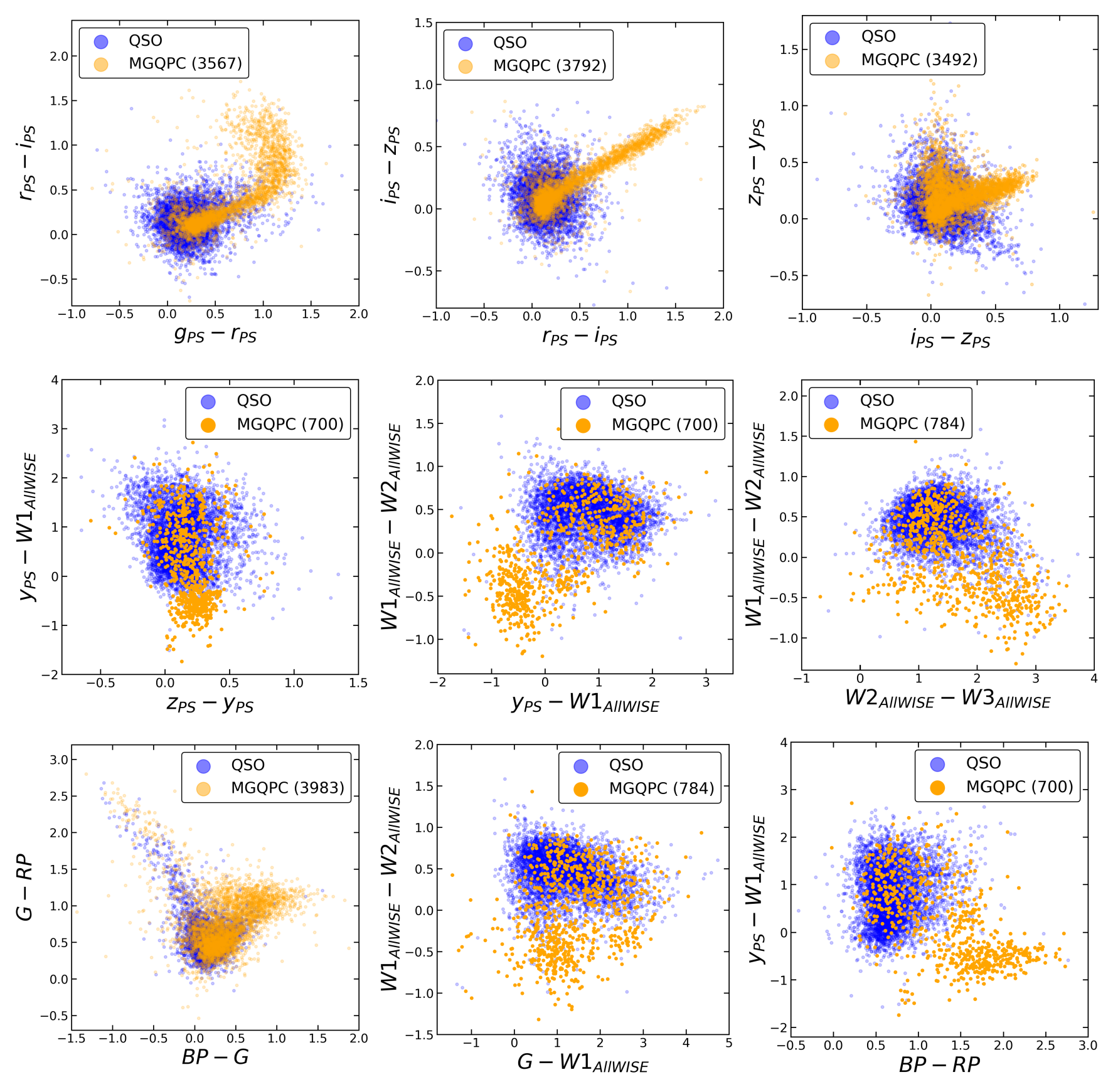}
    \caption{Color features of the MGQPCs. The number of the MGQPCs in each color diagram is consistently smaller than the total number of MGQPCs due to missing multiband photometry for some candidates, while the number of QSOs is equal to 5\,000 in each diagram. All magnitudes are given in the AB system.}
    \label{MGQPC_colors}
\end{figure*}

To evaluate the color properties of the MGQPCs, a subset of 5\,000 quasars was randomly selected from \texttt{GoodQSO\_Mag}. Nine types of color-color diagrams \citep[e.g.,][]{SF2024Quaia, FuYM2021GPQ-I, FuYM2024CatNorth, FuYM2025CatSouth} were applied, which incorporate multiband photometric data from PanSTARRS ($grizy$), AllWISE ($W1$, $W2$, and $W3$), and \textit{Gaia} ($G$, $BP$, and $RP$). The distributions of the 5\,000 known quasars and the MGQPCs were overlaid on these diagrams (Figure \ref{MGQPC_colors}). It should be noted that the number of MGQPCs varies across different color diagrams and is consistently smaller than the total number of MGQPCs due to missing multiband photometry for some candidates. In particular, diagrams involving AllWISE photometry contain only 700 to 800 MGQPCs. Nevertheless, they are still overlaid to partially reflect the color similarities or differences between the MGQPCs and known quasars. The color distribution of MGQPCs not only partially overlaps with known quasars but also exhibits stellar characteristics, such as the extended stripes and loci in the top row of the PanSTARRS color diagrams and obvious overdensities in other diagrams that do not coincide with quasar regions. Given the quasar selection efficiency of less than 70\,\% obtained in Section \ref{sec2.2.2}, these distributions can be well explained. However, candidates that do not overlap with the quasar distribution could include potential new quasars disguised by stellar colors. Such quasars with anomalous colors are often missed by large surveys (e.g., SDSS) during target selection, while astrometric quasar selection can effectively retrieve them. Therefore, we argue that retaining them for subsequent spectroscopic follow-up is necessary.

\subsection{Advantages of the MGQPC catalog} \label{sec3.2}
To clarify the sample advantages of the MGQPCs, three recently published and relatively larger candidate quasar pair catalogs or catalogs containing a substantial number of quasar pair candidates are introduced here for comparison with the MGQPC catalog at three aspects (member separation, \textit{Gaia} $G$-band magnitude, and redshift). The three catalogs are briefly described as follows:
\begin{enumerate}
    \item A catalog of 436 multiply lensed and binary quasar candidates (875 individual sources) obtained by \citet[][hereafter, Dawes23]{Dawes2023MultiLeQBQinDESILS} from the DESI-LS catalog by applying an autocorrelation algorithm.
    \item A catalog of lensed quasar and quasar pair candidates, comprising 971 systems (1\,977 individual sources), compiled by \citet[][hereafter, He23]{HeZZ2023DESILScatLeQ} based on the DESI-LS catalog and color similarity method.
    \item A golden sample of 1\,867 dual quasar candidates derived by \citet[][hereafter, Wu24]{WuQQ2024GMP-DULAG} using the GMP technique. 
\end{enumerate}

All of these catalogs were recompiled in this work by cross-matching SIMBAD to update the redshifts and remove the newly confirmed pairs, as well as by cross-matching \textit{Gaia} DR3 to retrieve the \textit{Gaia} coordinates and $G$-band magnitudes of their members. The member separations were calculated according to the member coordinates. We note that the candidates in Wu24 are not pairs of \textit{Gaia} sources, but single entries in \textit{Gaia} DR3 with possible multiple components and most of them have spectroscopic or photometric redshifts, with only 298 candidates lacking a redshift determination. After our recompilation, there were 412, 994, and 1\,867 candidate quasar pairs in Dawes23, He23, and Wu24, respectively. In Dawes23, there are 830 individual sources with \textit{Gaia} $G$-band magnitudes, while 150 systems have at least one member with spectroscopic redshifts. For He23, 1\,369 individual sources have \textit{Gaia} $G$-band magnitudes, while 348 candidate pairs have at least one member with spectroscopic redshifts. And for Wu24, all the candidates have \textit{Gaia} $G$-band magnitudes, while 1\,569 candidates have spectroscopic or photometric redshifts.

\begin{figure}
    \centering
    \includegraphics[width=0.48\textwidth]{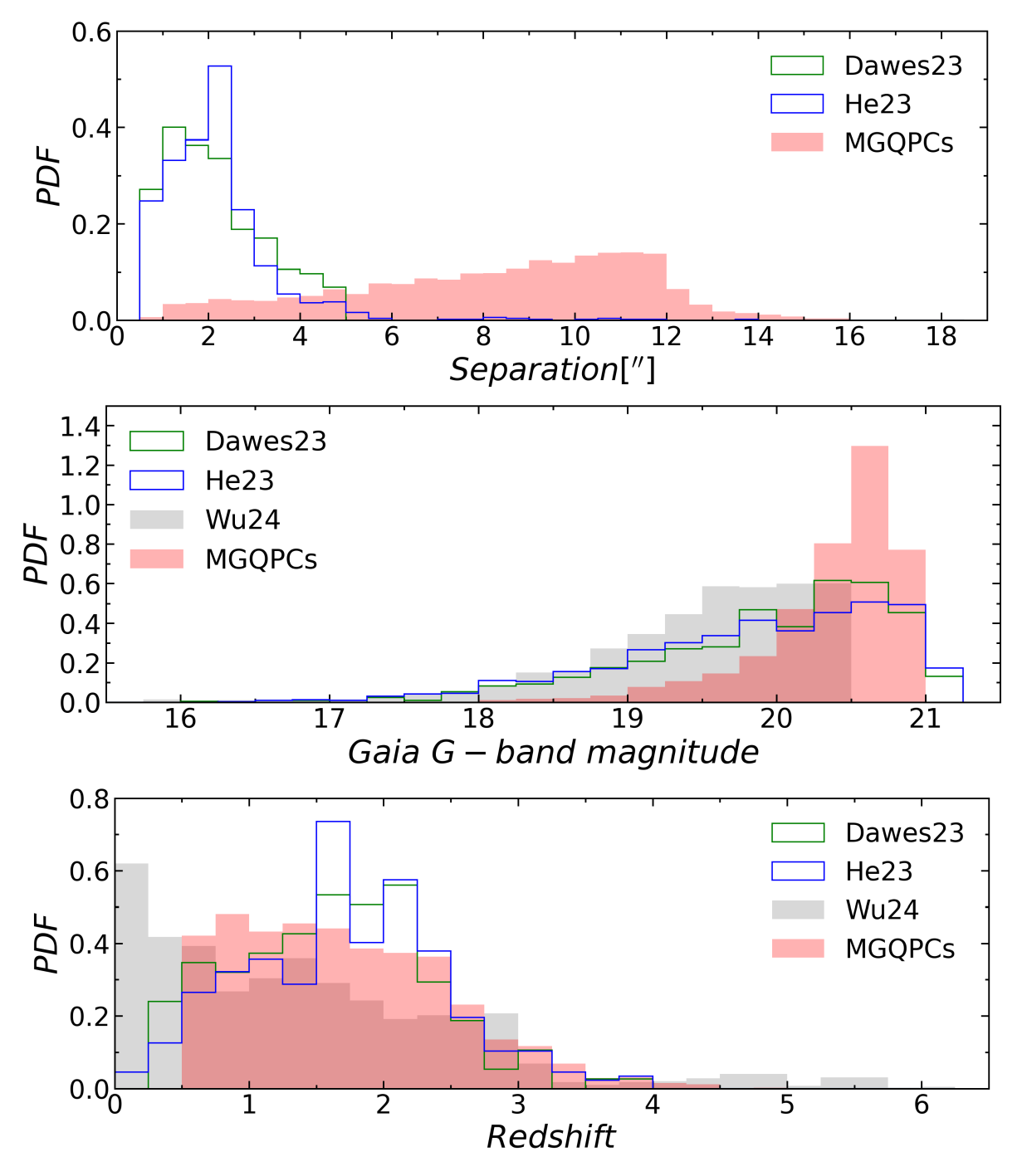}
    \caption{Comparison of Dawes23 (green unfilled), He23 (blue unfilled), Wu24 (gray filled), and the MGQPC catalog (red filled) on the distributions of the member separation, \textit{Gaia} $G$-band magnitude, and redshift. The y-axes are all scaled as probability distribution functions (PDF) to exhibit the distribution characteristics of the three parameters intuitively. Top panel: Separation histograms of Dawes23, He23, and the MGQPC catalog. Middle panel: \textit{Gaia} $G$-band magnitude histograms of the four catalogs. Bottom panel: Redshift histograms of the four catalogs.}
    \label{Comparison}
\end{figure}

The comparative histograms of member separation, \textit{Gaia} $G$-band magnitude, and redshift distributions are presented in Figure \ref{Comparison} for the three catalogs versus the MGQPCs. In terms of separation distribution (top panel), the Dawes23, He23, and MGQPCs cover ranges of 0.50{\arcsec}--4.99{\arcsec}, 0.50{\arcsec}--13.93{\arcsec}, and 0.66{\arcsec}--17.99{\arcsec}, respectively. The median separations of the three catalogs are 1.93{\arcsec}, 2.04{\arcsec}, and 8.81{\arcsec}, respectively. It should be noted that the separation distribution of Wu24 falls within \textit{Gaia}'s resolution range for detecting multipeak sources (0.15{\arcsec}--0.80{\arcsec}), making its exact distribution and median value uncertain; thus, its distribution is not plotted in the histogram. The substantial difference in separation distribution and the higher median value of the MGQPC catalog suggest it is more suitable for expanding the quasar pair sample to larger separations, especially in the 5{\arcsec}--18{\arcsec} range. For the magnitude distribution (middle panel), the Dawes23, He23, Wu24, and MGQPCs cover 16.15--21.60, 16.34--21.87, 14.96--20.50, and 16.34--21.00, respectively. Their median values are 20.12, 20.02, 19.66, and 20.52, indicating that the MGQPC catalog is slightly fainter and deeper than the other three catalogs. Finally, for the redshift distribution (bottom panel), the four catalogs cover 0.29--3.79, 0--3.80, 0--6.11, and 0.50--6.98, respectively. Their distribution characteristics are generally similar, and the median values are 1.67, 1.71, 1.25, and 1.61, respectively. However, there are still subtle differences: Dawes23 and He23 show higher fractions at z $\sim$ 1.5--2.5 than Wu24 and the MGQPC catalog, while Wu24 exhibits a significantly larger fraction at z $\lesssim$ 0.5 compared to the others. Furthermore, the MGQPC catalog shows marginally higher fractions at z $\sim$ 0.5--1.5 relative to the three catalogs. Moreover, compared to Wu24 (a dedicated dual quasar catalog), the MGQPC catalog contains significantly more sources at z $\sim$ 0.5--2.5. This suggests the MGQPC catalog might substantially augment more quasar pairs since cosmic noon \citep[e.g.,][]{ShenY2023QPfraction, Chen2023DQCosmicNoon, Gross2025VODKACosmicNoon}.

On the other hand, the MGQPCs were cross-matched with the three catalogs, yielding 87 matches in Dawes23, 115 in He23, and 11 in Wu24. After excluding these matched candidates (totaling 128 candidate systems), 3\,984 new candidates were obtained, which had not been included in Dawes23, He23, or Wu24. This demonstrates that the MGQPCs provide a significantly complementary update to existing major candidate quasar pair catalogs, particularly in terms of member separation coverage.

\subsection{Interesting candidates} \label{sec3.3}
The search for quasar pairs inevitably leads to the discovery of lensed quasars and vice versa \citep[e.g.,][]{Agnello2018LeQQP, Lemon2023GLQG-IV150LeQQP, YueMH2023HighZLeQQP, Dawes2023MultiLeQBQinDESILS, Makarov2023GaiaWISEDQLeQ, JiX2023SQUAB-IILeQDQsearch, JiX2024DQLeQinSpecSurveys}. Currently, most confirmed lensed quasars have small separations, with the majority being within 4{\arcsec}. However, several wide-separation lensed quasars \citep[WSLQs, e.g.,][]{Inada2003WSLQJ1004+4112, ShuYP2018WSLQJ0909+4449, Martinez2023WSLQJ0542-2125, Napier2023WSLQJ0335-1927} have also been discovered. These are typically the result of quasars being lensed by nearby galaxy groups or clusters. WSLQs can also occur when quasars are lensed by single galaxies, producing relatively large separations ranging from $\sim$ 6{\arcsec} to 10{\arcsec}. The largest separation of a WSLQ lensed by a single massive galaxy discovered to date is $\sim$ 10.1{\arcsec} \citep[J1651-0417, also known as Dragon Kite,][]{Ducourant2019GraLJ1651-0417DragonKite, Stern2021GraL-VIJ1651-0417DragonKite, Connor2022GraL-VIIJ1651-0417DragonKite}.

\begin{figure}
    \centering
    \includegraphics[width=0.48\textwidth]{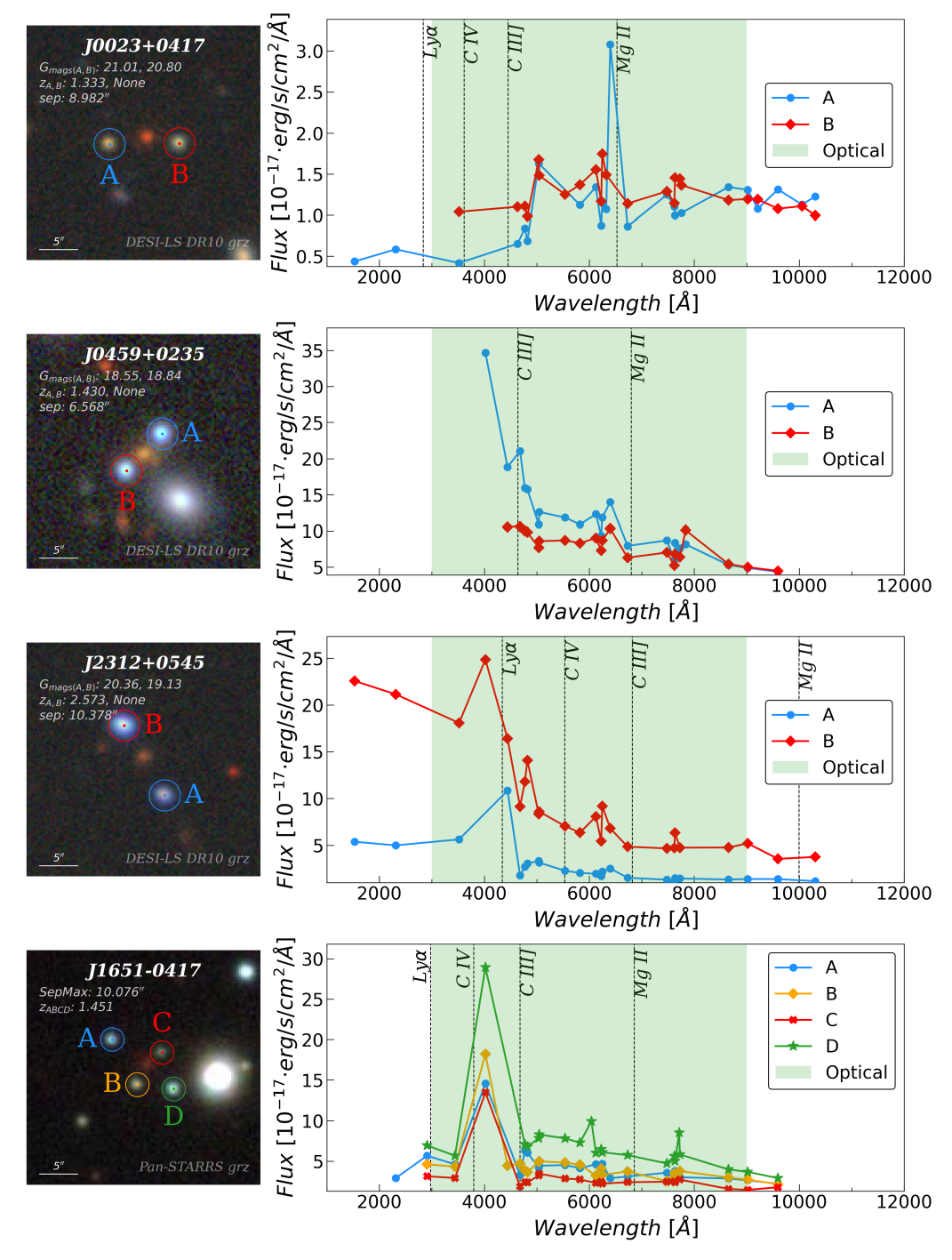}
    \caption{Three potential WSLQs or quasar-galaxy-quasar projection systems noted during the visual inspection, as well as the J1651-0417. Their SEDs show the similarity between the members in each pair system, with several typical emission lines of the known quasars (black dashed lines with line names) and an optical wavelength region of 3000{\AA}--9000{\AA} (green shaded area) overlapping on the SEDs. The marks and text annotations in the first three pseudo-color images are the same as in Figures \ref{DFs_NGs_examples} and \ref{MGQPC_examples}, while the four colored circles with center points, along with ``SepMax'' and ``$z_{ABCD}$'' denote the quadruply lensed quasars, the largest separation in the J1651-0417 (occurs between A and D), and the redshift of this lensed quasar, respectively.}
    \label{WSLQcands_examples}
\end{figure}

During the visual inspection, we made note of some quasar pair candidates. Their relatively large separation, similar color, and configuration indicate the possibility of being potential WSLQs. These WSLQ candidates exhibit separations from $\sim$ 6                                                                                                                                                                                                                                                                                                                                                                                                                                                                   {\arcsec} to 10{\arcsec}, with a potential galaxy located between them. Figure \ref{WSLQcands_examples} displays three of these potential WSLQs. For comparison, the unique WSLQ lensed by a single massive galaxy, J1651-0417, is also displayed in Figure \ref{WSLQcands_examples}. Their multiband photometric data (via SED) were retrieved from the VizieR database\footnote{\url{https://vizier.cds.unistra.fr/viz-bin/VizieR}}. Multi-epoch measurements of the same band (if any) were averaged to characterize the photometry of that band.

The SEDs of the candidate quasars are highly similar to those of known quasars paired with them. This would indicate similar redshifts and spectral features such as the continuum and/or broad emission lines. Together with the existence of possible foreground galaxies, these systems might be WSLQs lensed by massive galaxies. However, they may also be quasar-galaxy-quasar projection systems, which are precious natural laboratories of detailed study on the circumgalactic medium \citep[e.g.,][]{Krishnarao2022QuasarProbeLMC}.

\section{Discussion} \label{sec4}
This section discusses potential reasons for the insensitivity of the astrometric criteria to sky regions described in Section \ref{sec2.2.2}, along with the astrometric efficiency, the reason for avoiding additional color cuts to further improve selection efficiency, and an approximate evaluation of stellar contamination in crowded fields. Finally, several techniques to improve the success rate of quasar pair candidate selection are discussed.

\subsection{Quasar selection efficiency} \label{sec4.1}
The quasar selection purity, described in Section \ref{sec2.2.2}, remains invariant with Galactic latitude under identical astrometric criteria. Compared to Heintz20, we consider several potential reasons for this discrepancy. First, the astrometric criteria employed by Heintz20 differ slightly from ours: they utilized the signal-to-noise ratios of proper motion and parallax to characterize astrometric features; whereas we adopted the statistically more robust probability densities of zero proper motion and zero parallax. Second, Heintz20 relied on astrometric data from \textit{Gaia} DR2 to assess selection efficiency, while we utilized the latest \textit{Gaia} DR3, which may introduce subtle differences in the reference samples. Lastly, and most significantly, the selection of reference samples for quasars and stars differs. We constructed reference samples from the largest catalogs of spectroscopically confirmed quasars and stars that have \textit{Gaia} DR3 astrometry. These reference samples are substantially augmented with newly confirmed quasars from SDSS DR18 and DESI DR1, particularly at low and intermediate Galactic latitudes. We assert that the reference samples selected in Section \ref{sec2.2.2} exhibit higher completeness of sky distribution compared to those in Heintz20. Further validation may require a dedicated investigation.

Section \ref{sec2.2.2} reports comprehensive purity and stellar contamination of 69.37\,\% and 30.63\,\%, respectively. We consider this astrometric selection purity an upper limit, as the actual purity may be further reduced by several factors. Since galaxies were not included in the reference samples despite their astrometric similarities to quasars, their presence may also reduce the quasar selection purity. Although the visual inspection described in Section \ref{sec2.2.3} had removed some galaxies from the \texttt{MGQPCpreCAT}, we cannot unequivocally ensure the complete elimination of all galaxies. Furthermore, Figure \ref{mag_am_relation} and Table \ref{Tab_amcira_Gmag} demonstrate that magnitude significantly affects astrometric quasar selection, particularly for candidates with $G \geqslant$ 19. Therefore, the MGQPCs still contain approximately 31\,\% stellar contamination, which could be mitigated by incorporating additional quasar selection techniques such as color cuts.

The color diagrams of the MGQPCs indicate that some candidates exhibit apparent stellar color features (Figure \ref{MGQPC_colors}). These sources may be stellar contamination or could be quasars mimicking stellar colors. Such anomalous colors might result from interstellar and intergalactic reddening or interactions within quasar pairs. Applying color cuts may reduce the stellar contamination, but would also discard genuine quasars with unusual colors. Thus, we were able to preserve these candidates for subsequent follow-up confirmation. This is also the characteristic of this work: to search for quasars (and, hence, quasar pairs) that were missed by color selections.

\begin{figure*}
    \centering
    \includegraphics[width=0.9\textwidth]{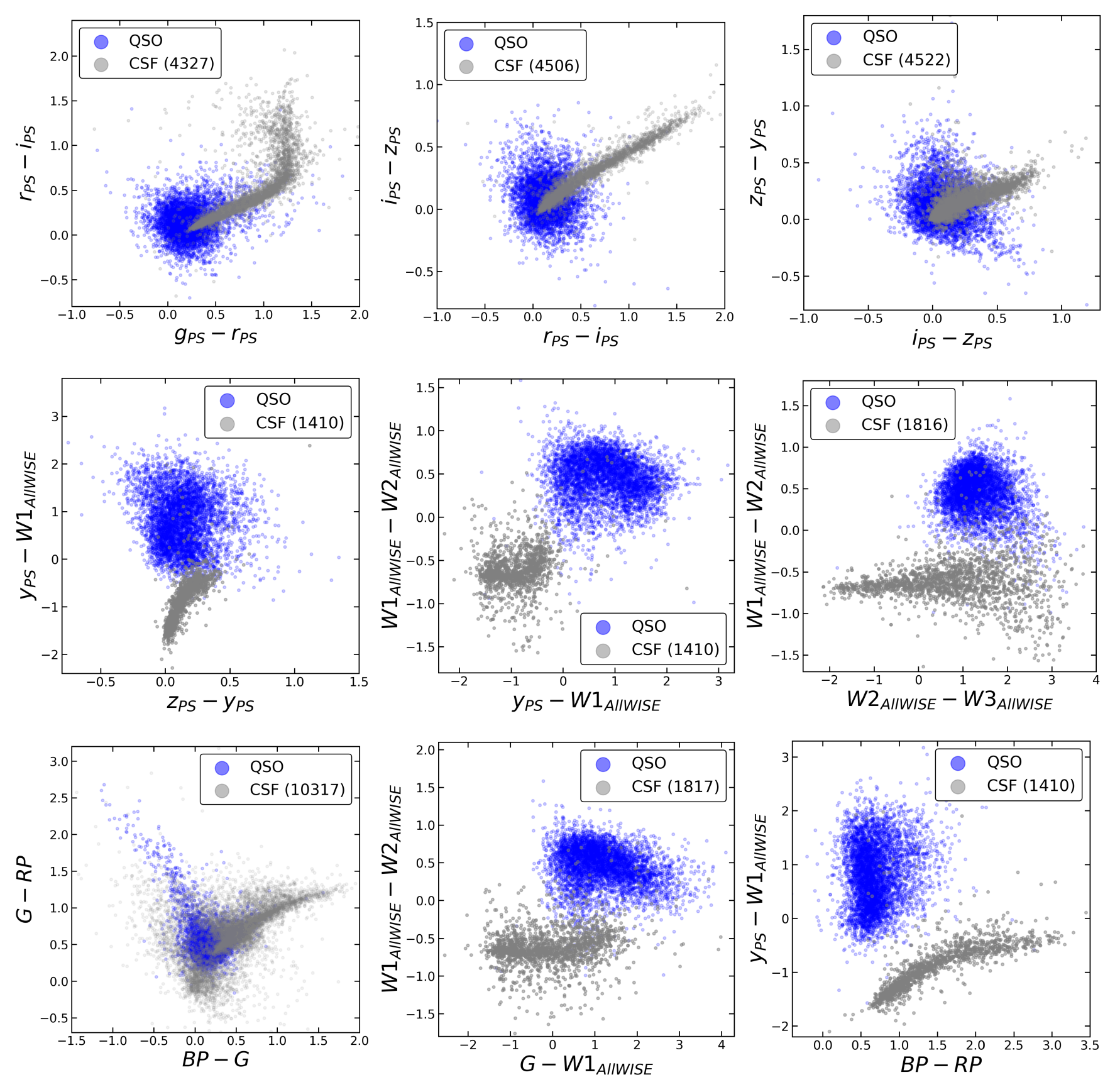}
    \caption{Color features of CSFs. The number of CSFs in each color diagram is consistently smaller than the total CSFs due to missing multiband photometry for some candidates, while the number of QSOs is equal to 5\,000 in each diagram. All magnitudes are in the AB system.}
    \label{DF_colors}
\end{figure*}

Since pure astrometric techniques are unsuitable for CSFs, we removed candidates in the three major CSFs (totaling 11,482 objects) from the \texttt{astrometricCAT} before applying astrometric filtering in Section \ref{sec2.2.2}. Here, we provide a brief assessment of stellar contamination for these candidates. Color diagrams identical to those used for the MGQPCs are presented in Figure \ref{DF_colors}. The distribution of CSF candidates aligns closely with that of stars, indicating severe stellar contamination. Thus, quasar selection in CSFs remains challenging, even combining astrometric filtering and color cuts. It would be even more difficult to search for quasar pairs in these regions.

\subsection{Future improvements} \label{sec4.2}
Although astrometric methods have been proven to be efficient in quasar searches in many studies, the search for quasar pairs not only requires the characteristic of zero proper motion and zero parallax, but also requires the redshifts of the two member quasars to be nearly identical. Therefore, more stringent criteria are still needed to refine the MGQPC catalog. To improve the success rate, several techniques might be helpful:

\begin{itemize}
    \item Estimating the photometric redshifts of the quasar candidates and selecting those with photometric redshifts similar to the spectroscopic redshifts of the nearby known quasars would be effective in selecting some high-confidence quasar pair candidates.
    \item Avoiding sky coverages with small stellar proper motions could make sense when searching for quasar candidates with astrometric criteria, as this can make quasars more distinguishable from stars and reduce the stellar contamination. In addition, avoiding sky coverages of CSFs could also reduce stellar contamination.
    \item Comparing the SEDs of the quasar candidates with those of the paired known quasars and selecting those pairs with high similarity in SEDs as high-probability quasar pair candidates would also work. We note that not all of their photometry across different bands was measured simultaneously, which would result in inaccurate SEDs. As a reference, and considering that the general variability of quasars is unlikely to change the features of their SEDs significantly, these photometric data can still be regarded as plausible and useful.
    \item In the color-based quasar selection, some quasars lying just outside the selection boundaries may be omitted. These quasars with slightly anomalous colors can be easily isolated through astrometric methods (several newly discovered quasars lying just outside the boundaries will be reported in a forthcoming paper). Thus, using the astrometric method would be an effective quasar selection method to isolate sources that are near, but still outside of the quasar selection boundaries.
    \item Since the fibers of existing spectroscopic surveys (e.g., SDSS, LAMOST, and DESI) have angular diameters of 1.5{\arcsec}--3{\arcsec}, those quasar pairs or lensed quasars with small separations ($\lesssim$ 3{\arcsec}) cannot be resolved spatially in fiber spectroscopy and they might be identified as single quasars. Therefore,  candidates with very small member separations can be set as high-priority.
\end{itemize}

Lastly, visual inspections can only remove nearby galaxies with apparent morphological characteristics, while removing high-redshift or morphologically uncharacteristic galaxies is difficult. Despite using morphological parameters from DESI-LS DR10, a small number of galaxies are still misclassified as point sources. Therefore, more precise morphological measurements are required to exclude galaxies with inactive nuclei and refine the quasar pair candidate sample.

\section{Summary} \label{sec5}
In this work, we utilized the current largest quasar collection, Million Quasar Catalog, to perform a cross-matching with \textit{Gaia} DR3 to systematically search for quasar pair candidates. The cross-matching radii were determined by a transverse distance of 100 kpc at the updated redshifts of the known quasars. Based on the reference samples of spectroscopically confirmed quasars and stars, multithreshold astrometric criteria based on the probability density were applied, isolating 4\,286 extragalactic candidates. A visual inspection was performed to remove CSFs and nearby galaxies. This resulted in 4\,112 candidates, which we designated as MGQPCs. Their median member separation is 8.81{\arcsec}, the \textit{Gaia} $G$-band median magnitude is 20.52, and the median redshift is 1.61. The MGQPC catalog was compared with three major candidate quasar pair catalogs (Dawes23, He23, and Wu24) to demonstrate its advantages and complementary roles in constructing the quasar pair population. After excluding 128 duplicates shared with the three catalogs, 3\,984 previously uncataloged candidates were identified. Some WSLQ candidates were picked out during visual inspection and are briefly described. We also discuss our quasar selection purity of approximately 69\,\%, as well as color properties of the MGQPCs and discarded candidates in the crowded fields. Several techniques for improving the success rate of quasar pair selection have been proposed, including photometric redshift estimation, sky coverage restriction, SED comparison, breaking of color methods, and unresolved quasar pairs in fiber spectroscopy. Approximately 31\,\% of the MGQPC catalog contains residual stellar contaminants, necessitating extensive spectroscopic follow-up observations for validation.

A spectroscopic follow-up using Xinglong 2.16m, Lijiang 2.4m, Palomar 200-inch telescope (P200), and LAMOST-III was recently conducted for several tens of MGQPCs. The details of their observations and identifications will be reported in a follow-up paper.

\section*{Data availability}
The MGQPC catalog is available at the CDS via \url{https://cdsarc.cds.unistra.fr/viz-bin/cat/J/A+A/707/A30}. Additionally, collaborative opportunities are open to conduct spectroscopic follow-up observations of the MGQPCs.

\bibliographystyle{aa}
\bibliography{aa55435-25}

\begin{appendix}
\section{Acknowledgments}
The authors thank the anonymous referee for the valuable comments that improved the quality and clarity of the manuscript. The authors thank Qirong Yuan, Heng Yu, He Gao, Yiping Shu, Xikai Shan, Yuming Fu, and Yuanzhen Han for their suggestions and guidance on several details. This work has been supported by the Chinese National Natural Science Foundation grant No. 12333001 and by the National Key R\&D Program of China (2021YFA0718500 and 2025YFA1614101).

This work has made use of data from the European Space Agency (ESA) mission \textit{Gaia} (\url{https://www.cosmos.esa.int/gaia}), processed by the \textit{Gaia} Data Processing and Analysis Consortium (DPAC, \url{https://www.cosmos.esa.int/web/gaia/dpac/consortium}). Funding for the DPAC has been provided by national institutions, in particular the institutions participating in the \textit{Gaia} Multilateral Agreement.

Funding for the Sloan Digital Sky Survey V has been provided by the Alfred P. Sloan Foundation, the Heising-Simons Foundation, the National Science Foundation, and the Participating Institutions. SDSS acknowledges support and resources from the Center for High-Performance Computing at the University of Utah. SDSS telescopes are located at Apache Point Observatory, funded by the Astrophysical Research Consortium and operated by New Mexico State University, and at Las Campanas Observatory, operated by the Carnegie Institution for Science. The SDSS website is \url{www.sdss.org}.

Guoshoujing Telescope (the Large Sky Area Multi-Object Fiber Spectroscopic Telescope, LAMOST) is a National Major Scientific Project built by the Chinese Academy of Sciences. Funding for the project has been provided by the National Development and Reform Commission. LAMOST is operated and managed by the National Astronomical Observatories, Chinese Academy of Sciences.

This research used data obtained with the Dark Energy Spectroscopic Instrument (DESI). DESI construction and operations are managed by the Lawrence Berkeley National Laboratory. This material is based upon work supported by the U.S. Department of Energy, Office of Science, Office of High-Energy Physics, under Contract No. DE–AC02–05CH11231, and by the National Energy Research Scientific Computing Center, a DOE Office of Science User Facility under the same contract. Additional support for DESI was provided by the U.S. National Science Foundation (NSF), Division of Astronomical Sciences under Contract No. AST-0950945 to the NSF’s National Optical-Infrared Astronomy Research Laboratory; the Science and Technology Facilities Council of the United Kingdom; the Gordon and Betty Moore Foundation; the Heising-Simons Foundation; the French Alternative Energies and Atomic Energy Commission (CEA); the National Council of Humanities, Science and Technology of Mexico (CONAHCYT); the Ministry of Science and Innovation of Spain (MICINN), and by the DESI Member Institutions: www.desi.lbl.gov/collaborating-institutions. The DESI collaboration is honored to be permitted to conduct scientific research on I’oligam Du’ag (Kitt Peak), a mountain with particular significance to the Tohono O’odham Nation. Any opinions, findings, and conclusions or recommendations expressed in this material are those of the author(s) and do not necessarily reflect the views of the U.S. National Science Foundation, the U.S. Department of Energy, or any of the listed funding agencies.

The DESI Legacy Imaging Surveys consist of three individual and complementary projects: the Dark Energy Camera Legacy Survey (DECaLS), the Beijing-Arizona Sky Survey (BASS), and the Mayall z-band Legacy Survey (MzLS). Pipeline processing and analyses of the data were supported by NOIRLab and the Lawrence Berkeley National Laboratory (LBNL). Legacy Surveys was supported by: the Director, Office of Science, Office of High Energy Physics of the U.S. Department of Energy; the National Energy Research Scientific Computing Center, a DOE Office of Science User Facility; the U.S. National Science Foundation, Division of Astronomical Sciences; the National Astronomical Observatories of China, the Chinese Academy of Sciences and the Chinese National Natural Science Foundation. LBNL is managed by the Regents of the University of California under contract to the U.S. Department of Energy.

The Pan-STARRS1 Surveys (PS1) and the PS1 public science archive have been made possible through contributions by the Institute for Astronomy, the University of Hawaii, the Pan-STARRS Project Office, the Max-Planck Society and its participating institutes, the Max Planck Institute for Astronomy, Heidelberg and the Max Planck Institute for Extraterrestrial Physics, Garching, The Johns Hopkins University, Durham University, the University of Edinburgh, the Queen's University Belfast, the Harvard-Smithsonian Center for Astrophysics, the Las Cumbres Observatory Global Telescope Network Incorporated, the National Central University of Taiwan, the Space Telescope Science Institute, the National Aeronautics and Space Administration under Grant No. NNX08AR22G issued through the Planetary Science Division of the NASA Science Mission Directorate, the National Science Foundation Grant No. AST–1238877, the University of Maryland, Eotvos Lorand University (ELTE), the Los Alamos National Laboratory, and the Gordon and Betty Moore Foundation.

We acknowledge the use of public data from the following facilities: \textit{Gaia}, SDSS, LAMOST, DESI, DESI Legacy Survey, and Pan-STARRS. This research made use of the following software packages: Astropy \citep{AstropyCollaboration2013, AstropyCollaboration2018, AstropyCollaboration2022}, Matplotlib \citep{Hunter2007Matplotlib}, NumPy \citep{Walt2011NumPy, Harris2020NumPy}, Pandas \citep{McKinney2010Pandas, pandas2022}, TOPCAT \citep{Taylor2005TOPCAT}, HumVI \citep{Marshall2015SWHumVI, Marshall2016HumVI}.
\end{appendix}

\end{document}